\documentclass[aps,prb,twocolumn,showpacs,amsmath,amssymb]{revtex4}
\usepackage{graphicx}
\usepackage{Jonasmacros}
\usepackage{bm}

\begin{document}
\title{Non-equilibrium theory for strongly coupled quantum dot with arbitrary on-site correlation strength}
\author{J. Fransson}
\email{Jonas.Fransson@fysik.uu.se}
\affiliation{Department of Materials Science and Engineering, Royal Institute of Technology (KTH), SE-100 44\ \ Stockholm, Sweden}
\affiliation{Physics Department, Uppsala University, Box 530, SE-751 21\ \ Uppsala, Sweden}
\affiliation{NORDITA, Blegdamsvej 17, DK-2100 Copenhagen, Denmark}

\begin{abstract}
An analytical expression for the current through a single level quantum dot for arbitrary strength of the on-site electron-electron interaction is derived beyond standard mean-field theory. By describing the localised states in terms of many-body operators, the employed diagrammatic technique for strong coupling enables inclusion of electron correlation effects into the description of the local dynamics, which provides transport properties that are consistent with recent experimental data.
\end{abstract}
\maketitle

\section{Introduction}
\label{sec-introduction}
Electron transport through few electron islands is strongly influenced by electron correlations between its atomic like states. A variety of correlation effects, including Kondo effect\cite{goldhaber1998,cronenwett1998,ji2000} and resonant current peaks,\cite{webb1985,johnson1992} have been observed. For low conductance of the tunnel junctions one can employ "orthodox theory", \cite{glazman1989,averin1991} which treats the tunnelling in lowest order perturbation theory (golden rule). In the strong coupling regime, however, this theory collapse since then the transport is not dominated by sequential tunnelling. While effects of coherent tunnelling have been extensively studied,\cite{aguado2000,brandes2000,konig2001,foa2003,abanin2004} the question of a general description which includes correlation effects and is valid both for strong and weak coupling regimes remains open.

In this paper, an analytical formula for the current through a single level quantum dot (QD) is derived for arbitrary on-site correlation strength, beyond standard mean-field theory (e.g. beyond self-consistent Hartree-Fock, or Hubbard I, approximation (HIA)\cite{hubbard,hewson1966,varma1976}). This formula  include local correlation effects which provide the well known renormalisation of the localised level.\cite{barabanov1974,varma1976,faddeev1961,haldane1978,ruckenstein1989,izyumov1992,izyumov1994} The theory and formula presented are shown to be consistent with previous model results for interactions between localised and delocalised electrons, e.g. atomic limit, non-interacting limit (on-site Coulomb repulsion $U\rightarrow0$), and the strongly correlated limit $(U\rightarrow\infty)$. While these limits have been treated several times, and the last limit has been treated in detail concerning Kondo physics,\cite{meir1991,hershfield1991,wingreen1994,sivan1996,martinek2003,choi2004} it is important that the here presented material includes these limits. It is also important to note that the renormalisation of the localised level often discussed in scaling theory,\cite{barabanov1974,varma1976,faddeev1961,haldane1978,ruckenstein1989,izyumov1992,izyumov1994} is included into the present formulation.

Already in 1987 Larkin and Matveev\cite{larkin1987} suggested a simple formula for current through a single resonant level, 
\[
J\sim\int\frac{\Gamma^L\Gamma^R}{(\omega-\dote{0})^2+(\Gamma/2)^2}
	[f_L(\omega)-f_R(\omega)]d\omega.
\]
Here, $\Gamma^{L/R}$, $\Gamma^L+\Gamma^R=\Gamma$, defines the coupling between the resonant level $\dote{0}$ and the leads, whereas $f_{L/R}(\omega)=f(\omega-\mu_{L/R})$  is the Fermi function for the left/right $(L/R)$ lead at the chemical potential $\mu_{L/R}$. The formula provides a very simple description of the single level QD with no on-site interaction, i.e. $U\rightarrow0$. Since then, this description has been further generalised to arbitrary interactions in the interacting region,\cite{meir1992,jauho1994} however, given in terms of non-equilibrium Green functions of the localised levels. Although this formulation provides a general framework for mesoscopic quantum systems in non-equilibrium, it nevertheless lacks the transparency found in simple single resonant case. It is therefore motivated to derive an analytical formula for the current through a single level QD for arbitrary on-site correlation $U$, which is not explicitly given in terms of non-equilibrium GFs. Thereto, it is desired that this formula goes  beyond any standard mean-field theory, e.g. HIA, and include effects of electron correlations of the localised states. However, one should not expect that, for instance, the Kondo effect may be included into this formula, since the simplicity of the intended expression would be lost by an adequate treatment of non-equilibrium Kondo physics. Effects related to Kondo physics are hence omitted in this paper.

The outline of the paper is as follows. The general model used for calculations of the current is introduced in Sec. \ref{sec-model}, whereas the specific model used in this paper is discussed in Sec. \ref{sec-SLQD}. In Sec. \ref{sec-HIA} this model is discussed for the case of Hubbard's approximation. Then, in Sec. \ref{sec-mbGF}, the formulation of the problem is discussed in terms of many-body operators, where the HIA is re-derived as well as a more advanced approximation of the system (loop correction), in which effects from electron correlations are included. Here, also the formula for the current through the system is derived, both in the HIA and in the loop correction. Charge and current conservation of the discussed approximations is proved in Sec. \ref{sec-cons}, a few numerical examples are considered in Sec. \ref{sec-numerical}, and the paper is finally summarised in Sec. \ref{sec-sum}.

\section{Model}
\label{sec-model}
In many cases of modelling mesoscopic systems for transport, it is reasonable to regard the leads and interacting region as separate subsystems which interact via a tunnelling Hamiltonian $\Hamil_T$. Single or coupled QDs attached to leads are examples of often studied systems where such an approximation is appropriate, where the generic model is given by
\begin{equation}
\Hamil=\Hamil_L+\Hamil_R+\Hamil_\text{int}+\Hamil_T.
\label{eq-gmod}
\end{equation}
Since the properties of the interacting region, $\Hamil_\text{int}$, often governs the transport properties of the system as a whole, the lead Hamiltonians are normally modelled as simple non-interacting electron gases, e.g. $\Hamil_{L/R}=\sum_{k\sigma\in L/R}\leade{k}\cdagger{k}\c{k}$, where $\cdagger{k}\ (\c{k})$ creates (annihilates) an electron in the left/right ($L/R$) lead at the energy $\leade{k}$ and spin $\sigma=\up,\down$. The tunnelling interaction, $\Hamil_T$, accounts for the tunnelling of electrons between the leads and the interacting region and in the simplest case only single electron tunnelling is taken into account. Hence,
\begin{equation}
\Hamil_T=\sum_{kn\sigma}(v_{kn\sigma}\cdagger{k}\d{n\sigma}+H.c.),
\label{eq-HT}
\end{equation}
where $v_{k\sigma}$ is the hybridisation between the localised and de-localised electrons in the leads and interacting region, respectively, and $\ddagger{n\sigma}\ (\d{n\sigma})$ creates (annihilates) an electron at the $n$th level in the interacting region.

The model for the interacting region, $\Hamil_\text{int}$, may be of a more or less complicated structure. Nevertheless, it is normally possible to separate it into a single electron zero Hamiltonian term and an interacting term, i.e. $\Hamil_\text{int}=\Hamil_0+\Hamil_i$, where $\Hamil_0=\sum_{n\sigma}\dote{n\sigma}\ddagger{n\sigma}\d{n\sigma}$, and where $\Hamil_i$ includes electron-electron interactions that take place in the interacting region.

With these basic assumptions, along with the anti-commutation relations $\anticom{\c{k}}{\csdagger{k'\sigma'}}=\delta_{kk'}\delta_{\sigma\sigma'}$, $\anticom{\d{\sigma}}{\ddagger{\sigma'}}=\delta_{\sigma\sigma'}$, and all other equal to zero, it is possible to derive an exact expression for the stationary current through the system\cite{jauho1994}
\begin{eqnarray}
J&=&-\frac{e}{2h}\tr\im\int\biggl\{[\bfGamma^L-\bfGamma^R]\bfG^<(\omega)
\nonumber\\&&
	+[f_L(\omega)\bfGamma^L-f_R(\omega)\bfGamma^R]
		[\bfG^r(\omega)-\bfG^a(\omega)]d\omega.
\label{eq-J}
\end{eqnarray}
Here, $\Gamma_{nn'\sigma}^{L/R}=2\pi\sum_{k\in L/R}v_{kn\sigma}^*v_{kn'\sigma}\delta(\omega-\leade{k})$ is the coupling between the leads and the interacting region, whereas $\bfG^{<,r,a}(\omega)$ are the lesser, retarded, and advanced forms of the GF matrix for the interacting region.

The model for the current given Eq. (\ref{eq-J}) is the basic equation for the calculations of transport properties in this paper. In the rest of the paper, the GF of the interacting region will be investigated for single QD in different approximations.

\section{Single level quantum dot}
\label{sec-SLQD}
Consider a single level QD with on-site Coulomb repulsion $U$, e.g.
\begin{equation}
\Hamil_\text{int}=\sum_\sigma\dote{0}\ddagger{\sigma}\d{\sigma}
	+Un_\up n_\down,
\label{eq-SLQD}
\end{equation}
where $\dote{0}$ is the single electron level and $n_\sigma=\ddagger{\sigma}\d{\sigma}$. Introduce the GF ${\cal G}_\sigma(t,t')=\eqgr{\d{\sigma}(t)}{\ddagger{\sigma}(t')}$. The equation of motion for the operator $\d{\sigma}$ is given by
\begin{equation}
\biggl(i\ddt-\dote{0}\biggr)\d{\sigma}=Un_{\bar\sigma}\d{\sigma}
	+\sum_kv_{k\sigma}^*\c{k},
\label{eq-eqmd}
\end{equation}
where $\bar\sigma$ is the opposite spin of $\sigma$. Further, the equations for $n_{\bar\sigma}\d{\sigma}$ and $\c{k}$ are given by
\begin{eqnarray}
\biggl(i\ddt-\dote{0}-U\biggr)n_{\bar\sigma}\d{\sigma}=
	\sum_k(-v_{k\bar\sigma}\d{\bar\sigma}\d{\sigma}\csdagger{k\bar\sigma}
\nonumber\\
			+v_{k\sigma}^*n_{\bar\sigma}\c{k}
			+v_{k\bar\sigma}^*\d{\sigma}\ddagger{\bar\sigma}\cs{k\bar\sigma}),
\label{eq-eqmnd}
\end{eqnarray}
and
\begin{equation}
\biggl(i\ddt-\leade{k}\biggr)\c{k}=v_{k\sigma}\d{\sigma},
\label{eq-eqmc}
\end{equation}
respectively. These equations lead to a coupled system including both one- and two-electron GFs, like for instance $\eqgr{n_{\bar\sigma}(t)\d{\sigma}(t)}{\ddagger{\sigma}(t')}$, $\eqgr{\c{k}(t)}{\ddagger{\sigma}(t')}$, etc. In order to proceed analytically (and/or numerically), one has to resort to one or another approximation. This will be done in the next subsection. However, the aim of this paper is to establish the equivalence between the described method and the method of using many-body (Hubbard) operators. Thus, before embarking into the details of the approximation schemes, the concept of using Hubbard operators will be introduced.

As is well known, the Fermi operators $\d{\sigma}\ (\ddagger{\sigma})$ can be expanded in the eigenstates of the interacting region. In the present case this is done by introducing the empty, singly, and doubly occupied states, e.g. $\ket{0}$, $\ket{\sigma}$, and $\ket{2}=\ket{\up\down}$, and the outer products (projection operators) $\X{pq}{}=\ket{p}\bra{q}$.\cite{hubbard} By resolution of unity one has
\begin{eqnarray}
\d{\sigma}&=&\sum_{pq}\ket{p}\bra{p}\d{\sigma}\ket{q}\bra{q}
	=\sum_{pq}\bra{p}\d{\sigma}\ket{q}\X{pq}{}
\nonumber\\
	&=&\bra{0}\d{\sigma}\ket{\sigma}\X{0\sigma}{}
		+\bra{\bar\sigma}\d{\sigma}\ket{2}\X{\bar\sigma2}{},
\label{eq-dexp}
\end{eqnarray}
and $\ddagger{\sigma}=\bra{\sigma}\d{\sigma}\ket{0}\X{\sigma0}{}+\bra{2}\d{\sigma}\ket{\bar\sigma}\X{2\bar\sigma}{}$, giving $\ddagger{\sigma}\d{\sigma}=\X{\sigma\sigma}{}+\X{22}{}$ and $n_\up n_\down=\X{22}{}$. Henceforth, Fermi-like transitions (changing the total number of electrons by an odd integer) will be denoted by $\X{pq}{}$, whereas Bose-like transitions (changing the total number of electrons by an even integer) are denoted by $\Z{pq}{}$, while diagonal transitions are denoted by $\h{p}{}=\Z{pp}{}$. The single level QD can then be written as 
\begin{equation}
\Hamil_\text{int}=\sum_{p=0,\sigma,2}E_p\h{p}{},
\label{eq-SLQDX}
\end{equation}
where $E_0=0$, $E_\sigma=\dote{0}$, and $E_2=2\dote{0}+U$.

In terms of the Hubbard operators, the tunnelling Hamiltonian takes the form
\begin{equation}
\Hamil_T=\sum_{k\sigma}(v_{k\sigma}\cdagger{k}
	[\X{0\sigma}{}+\eta_\sigma\X{\bar\sigma2}{}]+H.c),
\label{eq-HTX}
\end{equation}
where $\eta_\sigma=\bra{\bar\sigma}\d{\sigma}\ket{2}$ ($\eta_{\up,\down}=\pm1$) and $\bra{0}\d{\sigma}\ket{\sigma}=1$ account for the selection rules between the different transitions. In a similar way, the GF for the QD can be expanded according to
\begin{eqnarray}
\lefteqn{
{\cal G}_{\sigma}(t,t')=}
\nonumber\\&&=
	\noneqgrU{(\X{0\sigma}{}+\eta_\sigma\X{\bar\sigma2}{})(t)}
		{(\X{\sigma0}{}+\eta_\sigma\X{2\bar\sigma}{})(t')}
\nonumber\\&&=
	G_{0\sigma}(t,t')+\eta_\sigma G_{\bar\sigma2\sigma0}(t,t')
\nonumber\\&&\hspace{1cm}
		+\eta_\sigma G_{0\sigma2\bar\sigma}(t,t')
		+G_{\bar\sigma2}(t,t'),
\label{eq-Gdef}
\end{eqnarray}
(the subscript $U$ signifies the dependence of the generating functional, to be used later) where $G_{0\sigma}\equiv G_{0\sigma\sigma0}$ and $G_{\bar\sigma2}\equiv G_{\bar\sigma22\bar\sigma}$. In this example, the generating functional is defined by the action $S=\exp{[-i\kbint\Hamil'(t)dt]}$,\cite{sandalov2003,franssonPRL2002,kadanoff1963} where the disturbance potential is given by
\begin{equation}
\Hamil'(t)=U_0(t)\h{0}{}+\sum_{\sigma\sigma'}U_{\sigma\sigma'}(t)\Z{\sigma\sigma'}{}
	+U_2(t)\h{2}{}.
\label{eq-dist}
\end{equation}

The equation of motion for the operators $\X{0\sigma}{}$ and $\X{\bar\sigma2}{}$ are given by
\begin{eqnarray}
\biggl(i\ddt-\Delta_{\sigma0}^0\biggr)\X{0\sigma}{}=
	\sum_k(-\eta_{\bar\sigma}v_{k\bar\sigma}\csdagger{k\bar\sigma}\Z{02}{}
\nonumber\\
		+v_{k\sigma}^*(\h{0}{}+\h{\sigma}{})\c{k}
		+v_{k\bar\sigma}^*\Z{\bar\sigma\sigma}{}\cs{k\bar\sigma}),
\label{eq-eqmX0}
\end{eqnarray}
and
\begin{eqnarray}
\biggl(i\ddt-\Delta_{2\bar\sigma}^0\biggr)\X{\bar\sigma2}{}=
	\sum_k(-v_{k\bar\sigma}\csdagger{k\bar\sigma}\Z{02}{}
\nonumber\\
		+\eta_\sigma v_{k\sigma}^*(\h{\bar\sigma}{}+\h{2}{})\c{k}
		+\eta_{\bar\sigma}v_{k\bar\sigma}^*\Z{\bar\sigma\sigma}{}\cs{k\bar\sigma}),
\label{eq-eqmX0}
\end{eqnarray}
where $\Delta_{\sigma0}^0=E_\sigma-E_0$ and $\Delta_{2\bar\sigma}^0=E_2-E_{\bar\sigma}$ are the bare transition energies in the interacting region. In the following treatment, terms like $\csdagger{k\bar\sigma}\Z{02}{}$ are neglected. These terms give rise to propagators similar to those considered in the theory of superconductivity,\cite{sandalov2003,abrikosov1963} which is beyond the scope of the present paper.

\section{Hubbard's approximation}
\label{sec-HIA}
One of the simplest approximations of the GF for the QD is obtained within Hubbard's approximation of the two-electron GF, henceforth referred to as the Hubbard I approximation (HIA). For completeness, the result is derived in both the Fermi operator and the Hubbard operator representations, in order to elucidate the correspondence between the two approaches. The equation of motion for the two-electron GF ${\cal G}_{\sigma\bar\sigma}(t,t')=\eqgr{n_{\bar\sigma}(t)\d{\sigma}(t)}{\ddagger{\sigma}(t)}$ is given by, c.f. Eq. (\ref{eq-eqmnd}),
\begin{eqnarray*}
\biggl(i\ddt-\dote{0}-U\biggr){\cal G}_{\sigma\bar\sigma}(t,t')=
	\delta(t-t')\av{n_{\bar\sigma}(t)}
\\
	-\sum_kv_{k\bar\sigma}
		\eqgr{(\d{\bar\sigma}\d{\sigma}\csdagger{k\bar\sigma})(t)}
				{\ddagger{\sigma}(t')}
\\
	+\sum_kv_{k\sigma}^*\eqgr{(n_{\bar\sigma}\c{k})(t)}{\ddagger{\sigma}(t')}
\\
	+\sum_kv_{k\bar\sigma}^*
				\eqgr{(\d{\sigma}\ddagger{\bar\sigma}\cs{k\bar\sigma})(t)}
					{\ddagger{\sigma}(t')}.
\end{eqnarray*}
The HIA corresponds to the de-couplings\cite{varma1976}
\begin{eqnarray*}
\eqgr{(n_{\bar\sigma}\c{k})(t)}{\ddagger{\sigma}(t')}
	&=&\av{n_{\bar\sigma}(t)}F_{k\sigma}(t,t'),
\\
\eqgr{(\d{\bar\sigma}\d{\sigma}\csdagger{k\bar\sigma})(t)}
	{\ddagger{\sigma}(t')}&=&0,
\\
\eqgr{(\d{\sigma}\ddagger{\bar\sigma}\cs{k\bar\sigma})(t)}
		{\ddagger{\sigma}(t')}&=&0,
\end{eqnarray*}
where $F_{k\sigma}(t,t')=\eqgr{\c{k}(t)}{\ddagger{\sigma}(t')}$, which yields
\begin{eqnarray*}
\biggl(i\ddt-\dote{0}-U\biggr){\cal G}_{\sigma\bar\sigma}(t,t')=
	\delta(t-t')\av{n_{\bar\sigma}(t)}
\nonumber\\
	+\sum_kv_{k\sigma}^*\av{n_{\bar\sigma}(t)}F_{k\sigma}(t,t').
\end{eqnarray*}
Using Eq. (\ref{eq-eqmc}), the equation for the transfer GF $F_{k\sigma}(t,t')$ can be integrated to
\begin{equation}
F_{k\sigma}(t,t')=v_{k\sigma}\kbint g_{k\sigma}(t,t''){\cal G}_\sigma(t'',t')dt'',
\label{eq-F}
\end{equation}
where $g_{k\sigma}(t,t')=\eqgr{\c{k}(t)}{\cdagger{k}(t')}$ is the GF for the electrons in the leads satisfying $(i\ddtinline-\leade{k})g_{k\sigma}(t,t')=\delta(t-t')$. Hence, the (Fourier transformed) QD GF ${\cal G}_{\sigma}(t,t')$ reduces to the well known result\cite{hewson1966}
\begin{equation}
{\cal G}_\sigma(i\omega)=
	\frac{i\omega-\dote{0}-(1-\av{n_{\bar\sigma}})U}
		{[i\omega-\dote{0}-V_\sigma][i\omega-\dote{0}-U]
			-\av{n_{\bar\sigma}}UV_\sigma},
\label{eq-GHIA}
\end{equation}
where $V_\sigma\equiv V_\sigma(i\omega)=\sum_k|v_{k\sigma}|^2/(i\omega-\leade{k})$.

The HIA for arbitrary $U$ contains three important limit results for, namely, the atomic limit $v_{k\sigma}\rightarrow0$, the non-interacting limit $U\rightarrow0$ and the strongly correlated limit $U\rightarrow\infty$, e.g.
\begin{equation}
\lim_{v_{k\sigma}\rightarrow0}{\cal G}_\sigma(i\omega)=
	\frac{1-\av{n_{\bar\sigma}}}{i\omega-\dote{0}}+\frac{\av{n_{\bar\sigma}}}{i\omega-\dote{0}-U},
\label{eq-atomic}
\end{equation}
\begin{equation}
\lim_{U\rightarrow0}{\cal G}_\sigma(i\omega)=
	\frac{1}{i\omega-\dote{0}-V_\sigma(i\omega)},
\label{eq-HIA0U}
\end{equation}
e.g. the result from the exactly solvable Fano-Anderson model,\cite{fano1961,anderson1961} and\cite{varma1976}
\begin{equation}
\lim_{U\rightarrow\infty}{\cal G}_\sigma(i\omega)=
	\frac{1-\av{n_{\bar\sigma}}}
		{i\omega-\dote{0}-(1-\av{n_{\bar\sigma}})V_\sigma(i\omega)},
\label{eq-HIAinfU}
\end{equation}
respectively. The results in Eqs. (\ref{eq-GHIA}), (\ref{eq-atomic}), (\ref{eq-HIA0U}), and (\ref{eq-HIAinfU}) will now serve as a basis to compare the analogous result obtained within the approach with Hubbard operators.

\section{Formulation in terms of many-body operators}
\label{sec-mbGF}
The expansion of the operator $\d{\sigma}$ in terms of Hubbard operators, Eq. (\ref{eq-dexp}), provides the basic rules of the expansion of the QD GF ${\cal G}_\sigma$ given in Eq. (\ref{eq-Gdef}). In the latter expansion the propagators are constructed to depend on the generating functional, Eq. (\ref{eq-dist}), through the definition
\begin{eqnarray*}
G_{0\sigma}(t,t')&=&\noneqgrU{\X{0\sigma}{}(t)}{\X{\sigma0}{}(t')}\equiv
\\
	&\equiv&\noneqgr{\X{0\sigma}{}(t)}{\X{\sigma0}{}(t')},
\end{eqnarray*}
and similarly for the other GFs. In taking the time derivative of the GF with respect to $t$ one must also differentiate $S$, i.e. (c.f. Ref. \onlinecite{kadanoff1963})
\begin{eqnarray*}
i\ddt{\rm T}S\X{0\sigma}{}(t)&=&{\rm T}\Biggl[\biggl(i\ddt S\biggr)\X{0\sigma}{}(t)
		+S\biggl(i\ddt\X{0\sigma}{}(t)\biggr)\Biggr]
\\
	&=&{\rm T}S\com{\X{0\sigma}{}(t)}{\Hamil+\Hamil'(t)}.
\end{eqnarray*}
With the disturbance potential given in Eq. (\ref{eq-dist}) the commutator $\com{\X{a}{}}{\Hamil'}$, $a=0\sigma$ or $\bar\sigma2$, becomes
\begin{eqnarray*}
\com{\X{0\sigma}{}}{\Hamil'}&=&
	\Delta U_{\sigma0}(t)\X{0\sigma}{}+U_{\sigma\bar\sigma}(t)\X{0\bar\sigma}{},
\\
\com{\X{\bar\sigma2}{}}{\Hamil'}&=&
	\Delta U_{2\bar\sigma}(t)\X{\bar\sigma2}{}-U_{\sigma\bar\sigma}(t)\X{\sigma2}{},
\end{eqnarray*}
where $\Delta U_{\sigma0}(t)=U_{\sigma\sigma}(t)-U_0(t)$ and $\Delta U_{2\bar\sigma}(t)=U_2(t)-U_{\bar\sigma}(t)$. Hence, omitting transitions like $\Z{02}{}$, the equations of motion for the GFs $G_{0\sigma\bar{a}}(t,t')$ and $G_{\bar\sigma2\bar{a}}(t,t')$, where $\bar{a}$ denotes the conjugate of any of the transitions $0\sigma$ and $\bar\sigma2$, become
\begin{subequations}
\label{eq-G}
\begin{eqnarray}
\lefteqn{
\biggl(i\ddt-\Delta_{\sigma0}^0-\Delta U_{\sigma0}(t)\biggr)G_{0\sigma\bar{a}}(t,t')
}
\nonumber\\&&\vphantom{\biggl(}
	-U_{\sigma\bar\sigma}(t)G_{0\bar\sigma\bar{a}}(t,t')=
	\delta(t-t')P_{0\sigma\bar{a}}(t)
\nonumber\\&&
		+\sum_{k}\biggl(
		v_{k\sigma}^*
			\noneqgrU{([\h{0}{}+\h{\sigma}{}]\c{k}(t)}{\X{\bar{a}}{}(t')}
\nonumber\\&&
		+v_{k\bar\sigma}^*
			\noneqgrU{(\Z{\bar\sigma\sigma}{}\cs{k\bar\sigma})(t)}{\X{\bar{a}}{}(t')}
				\biggr)
\label{eq-G1}\\
\lefteqn{
\biggl(i\ddt-\Delta_{2\bar\sigma}^0
	-\Delta U_{2\bar\sigma}(t)\biggr)G_{\bar\sigma2\bar{a}}(t,t')
}
\nonumber\\&&\vphantom{\biggl(}
		+U_{\sigma\bar\sigma}(t)G_{\sigma2\bar{a}}(t,t')=
	\delta(t-t')P_{\bar\sigma2\bar{a}}(t)
\nonumber\\&&
		+\sum_k\biggl(
		\eta_\sigma v_{k\sigma}^*
			\noneqgrU{([\h{\bar\sigma}{}+\h{2}{}]\c{k}(t)}{\X{\bar{a}}{}(t')}
\nonumber\\&&
		+\eta_{\bar\sigma}v_{k\bar\sigma}^*
			\noneqgrU{(\Z{\bar\sigma\sigma}{}\cs{k\bar\sigma})(t)}{\X{\bar{a}}{}(t')}
				\biggr).
\label{eq-G2}
\end{eqnarray}
\end{subequations}
Here, 
$P_{0\sigma\bar{a}}(t)\equiv\occu{\anticom{\X{0\sigma}{}}{\X{\bar{a}}{}}(t)}$ and $P_{\bar\sigma2\bar{a}}(t)\equiv\occu{\anticom{\X{\bar\sigma2}{}}{\X{\bar{a}}{}}(t)}$ (\emph{end-factors}) are spectral weights of the respective GFs , playing an important role in this formulation of the theory. Also, let $P_{0\sigma}\equiv P_{0\sigma\sigma0}$ and $P_{\bar\sigma2}\equiv P_{\bar\sigma22\bar\sigma}$, for a shorter notation. Physical quantities are drawn out of the involved GFs in the limit $U_\xi(t)\rightarrow0$, since the sources are introduced in order to generate a diagrammatic expansion of the GFs only.

The present diagrammatic expansion of the GFs is generated through functional derivatives of the GFs with respect to the source fields $U_\xi(t)$ in Eq. (\ref{eq-dist}), where the functional differentiation operators arise from the three operator propagators $\noneqgrU{\Z{\xi}{}(t'')\c{k}(t)}{\X{\bar{b}}{}(t')}$. To see this, consider the variation of the GF, say, $G_{0\sigma}(t,t')$ with respect to the source fields $U_\xi(t)$, e.g. $\delta G_{0\sigma}(t,t')$. Through an analogous procedure as described in Ref. \onlinecite{kadanoff1963}, one finds that
\begin{eqnarray}
\lefteqn{
\noneqgrU{\Z{\xi}{}(t'')\X{0\sigma}{}(t)}{\X{\sigma0}{}(t')}=
}
\nonumber\\&&
=\biggl(\occu{\Z{\xi}{}(t'')}+i\varder{}{U_\xi(t'')}\biggr)G_{0\sigma}(t,t'),
\label{eq-varder}
\end{eqnarray}
and likewise for the other propagators. Hence, the equations of motion in Eq. (\ref{eq-G}) can be rewritten in terms of the normal two operator GFs and functional derivatives thereof. Using that the transfer GF $F_{k\sigma\sigma0}(t,t')=\noneqgrU{\c{k}(t)}{\X{\sigma0}{}(t')}$ can be integrated similarly as $F_{k\sigma}$ giving
\begin{eqnarray}
F_{k\sigma\sigma0}(t,t')=v_{k\sigma}\kbint g_{k\sigma}(t,t'')
	[G_{0\sigma}(t'',t')
\nonumber\\
	+\eta_\sigma G_{\bar\sigma2\sigma0}(t'',t')]dt''.
\label{eq-F0}
\end{eqnarray}
 and Eq. (\ref{eq-varder}), the equation of motion for the matrix GF can be written as
\begin{eqnarray}
\lefteqn{
\biggl(i\ddt I-\bfDelta^0-\bfU(t)\biggr)\bfG(t,t')=\delta(t-t')\bfP(t)
}
\nonumber\\&&
	+[\bfP(t^+)+\bfR(t^+)]\kbint\bfV(t,t'')\bfG(t'',t')dt'',
\label{eq-G'}
\end{eqnarray}
where $I$ is the identity, $\bfDelta^0=\mbox{diag}\{\Delta_{\up0}^0,\Delta_{\down0}^0,\Delta_{2\down}^0,\Delta_{2\up}^0\}$ (diagonal matrix) contains the bare transition energies $\Delta_{\sigma0}^0=E_\sigma-E_0$ and $\Delta_{2\bar\sigma}^0=E_2-E_{\bar\sigma}$. The source fields are contained in $\bfU(t)=\mbox{diag}\{\bfU_1(t),\bfU_2(t)\}$, where $\bfU_n(t),\ n=1,2$, are $2\times2$ matrices defined by
\begin{subequations}
\label{eq-U}
\begin{eqnarray}
\bfU_1(t)&=&\left(\begin{array}{cc}
	\Delta U_{\up0}(t) & U_{\up\down}(t) \\
	U_{\down\up}(t) & \Delta U_{\down0}(t)
	\end{array}\right)
\label{eq-U1}\\
\bfU_2(t)&=&\left(\begin{array}{cc}
	\Delta U_{2\down}(t) & -U_{\up\down}(t) \\
	-U_{\down\up}(t) & \Delta U_{2\up}(t)
	\end{array}\right).
\label{eq-U2}
\end{eqnarray}
\end{subequations}

The end-factor $\bfP(t)=\mbox{diag}\{\bfP_1(t),\bfP_2(t)\}$, which arise due to the non-commutativity of the Hubbard operators, contains the spectral weights of the components. Here, each entry $\bfP_n(t),\ n=1,2$, is given by
\begin{eqnarray*}
\bfP_1(t)&=&\left(\begin{array}{cc} 
	P_{0\up}(t) & P_{0\up\down0}(t) \\ 
	P_{0\down\up0}(t) & P_{0\down}(t) \end{array}\right),
\\
\bfP_2(t)&=&\left(\begin{array}{cc} 
	P_{\down2}(t) & P_{\down22\up}(t) \\ 
	P_{\up22\down}(t) & P_{\up22\up}(t) \end{array}\right),
\end{eqnarray*}
where the vanishing off-diagonal components are inserted for completness. It may be noted though, that the off-diagonal components in $\bfP_n,\ n=1,2$, are non-vanishing whenever spin-flip transitions occur in the system. In the spin degenerate case the system reduces to a $2\times2$ matrix equation, since the spin $\up$ and $\down$ equations are equal.

The functional differentiation operator matrix $\bfR=\mbox{diag}\{\bfR_1,\bfR_2\}$ has been introduced, arising by the same arguments as $\bfP(t)$, where $\bfR_n,\ n=1,2$ are defined by
\begin{eqnarray*}
\bfR_1(t)&=&\left(\begin{array}{cc}
	R_{0\up\up0}(t) & R_{0\up\down0}(t) \\
	R_{0\down\up0}(t) & R_{0\down\down0}(t)
	\end{array}\right),
\\
\bfR_2(t)&=&\left(\begin{array}{cc}
	R_{\down22\down}(t) & R_{\down22\up}(t) \\
	R_{\up22\down}(t) & R_{\up22\up}(t)
	\end{array}\right),
\end{eqnarray*}
with the components
\begin{subequations}
\label{eq-R}
\begin{equation}
R_{0\sigma\sigma'0}(t)=i\biggl(\delta_{\sigma\sigma'}\varder{}{U_0(t)}
	+\varder{}{U_{\sigma'\sigma}(t)}\biggr),
\label{eq-R0}
\end{equation}
\begin{equation}
R_{\bar\sigma22\bar\sigma'}(t)=i\biggl(\varder{}{U_{\bar\sigma\bar\sigma'}(t)}
	+\delta_{\bar\sigma\bar\sigma'}\varder{}{U_{2}(t)}\biggr).
\label{eq-R2}
\end{equation}
\end{subequations}
Note the order of the spin indices in the second term of Eq. (\ref{eq-R0}).
Finally, the tunnelling interaction matrix $\bfV$ is given by
\begin{equation}
\bfV(t,t')=\left(\begin{array}{cc}
	\bfV'(t,t') & \sigma_z\bfV'(t,t') \\
	\sigma_z\bfV'(t,t') & \bfV'(t,t')
	\end{array}\right),
\label{eq-MV}
\end{equation}
where $\bfV'(t,t')=\mbox{diag}\{V_\up(t,t'),V_\down(t,t')\}$, with $V_\sigma(t,t')=\sum_{k\in L,R}|v_{k\sigma}|^2g_{k\sigma}(t,t')$, and $\sigma_z$ is the $z$-component of the Pauli spin vector which accounts for the selection rules defined by $\eta_\sigma=\bra{\bar\sigma}\d{\sigma}\ket{2}$.

This concludes the definitions of the equations for the QD GF in terms of many-body operators in its general form. The next step is to find valuable approximations of the local properties that can be used in the non-equilibrium description of the system.

\subsection{Hubbard I approximation}
\label{ssec-mbHIA}
The HIA corresponds to omitting all functional derivatives, and in addition, putting all the averages $\occu{\Z{\xi}{}(t)}$ but the diagonal ones, $\occu{(\h{0}{}+\h{\sigma}{})(t)}$ and $\occu{(\h{\bar\sigma}{}+\h{2}{})(t)}$, equal to zero. Hence, the Fourier transformed equation of motion for $\bfG$ in the HIA becomes (as $U_\xi(t)\rightarrow0$)
\begin{equation}
(i\omega I-\bfDelta^0)\bfG(i\omega)=\bfP
	+\bfP\bfV(i\omega)\bfG(i\omega)
\label{eq-MHIA}
\end{equation}
which yields the solution
\[
G_{0\sigma}(i\omega)=\frac{P_{0\sigma}}
	{i\omega-\Delta_{\sigma0}^0-P_{0\sigma}V_\sigma
	-\frac{P_{0\sigma}V_\sigma P_{\bar\sigma2}}
		{i\omega-\Delta_{2\bar\sigma}^0-P_{\bar\sigma2}V_\sigma}V_\sigma},
\]
and
\[ G_{\bar\sigma2\sigma0}(i\omega)=\eta_\sigma 
	\frac{P_{\bar\sigma2}V_\sigma}
		{i\omega-\Delta_{2\bar\sigma}^0-P_{\bar\sigma2}V_\sigma}
			G_{0\sigma}(i\omega).
\]
The end-factors $P_{0\sigma},\ P_{\bar\sigma2}$, interpreted as spectral weights, have to add up to unity, i.e. $P_{0\sigma}+P_{\bar\sigma2}=1$ (proved in Sec. \ref{sec-cons}). For later use, one also notes that the occupation number $\av{n_\sigma}=\av{\h{\sigma}{}+\h{2}{}}=P_{\sigma2}$, hence $P_{0\sigma}=1-\av{n_{\bar\sigma}}$. By simple algebraic manipulations one thus finds that
\begin{equation}
G_{0\sigma}(i\omega)=\frac{(i\omega-\Delta_{2\bar\sigma}^0
	-P_{\bar\sigma2}V_\sigma)P_{0\sigma}}
		{[i\omega-\Delta_{\sigma0}^0-V_\sigma]
			[i\omega-\Delta_{2\bar\sigma}^0]-P_{\bar\sigma2}UV_\sigma}.
\label{eq-G0}
\end{equation}
Therefore, the sum $G_{0\sigma}+\eta_\sigma G_{\bar\sigma2\sigma0}$ gives the expression
\begin{eqnarray}
\lefteqn{
G_{0\sigma}(i\omega)+\eta_\sigma G_{\bar\sigma2\sigma0}(i\omega)=
}
\nonumber\\&&
=\frac{(i\omega-\Delta_{2\bar\sigma}^0)P_{0\sigma}}
	{[i\omega-\Delta_{\sigma0}^0-V_\sigma]
		[i\omega-\Delta_{2\bar\sigma}^0]-P_{\bar\sigma2}UV_\sigma}.
\label{eq-G01}
\end{eqnarray}
Similarly, one finds that the sum $\eta_\sigma G_{0\sigma2\bar\sigma}+G_{\bar\sigma2}$ can be written as
\begin{eqnarray}
\lefteqn{
\eta_\sigma G_{0\sigma2\bar\sigma}(i\omega)+G_{\bar\sigma2}(i\omega)=
}
\nonumber\\&&
=\frac{(i\omega-\Delta_{\sigma0}^0)P_{\bar\sigma2}}
	{[i\omega-\Delta_{\sigma0}^0-V_\sigma]
		[i\omega-\Delta_{2\bar\sigma}^0]-P_{\bar\sigma2}UV_\sigma}.
\label{eq-G21}
\end{eqnarray}
Thus, one finally arrives at (recalling that $\Delta_{\sigma0}^0=E_\sigma-E_0=\dote{0}$ and $\Delta_{2\bar\sigma}^0=E_2-E_\sigma=\dote{0}+U$),
\begin{eqnarray}
\vphantom{\frac{1}{1}}
G_{0\sigma}(i\omega)+\eta_\sigma G_{\bar\sigma2\sigma0}(i\omega)
	+\eta_\sigma G_{\sigma02\bar\sigma}(i\omega)+G_{\bar\sigma2}(i\omega)=
\nonumber\\
=\frac{(i\omega-\Delta_{\sigma0}^0)(P_{0\sigma}+P_{\bar\sigma2})-UP_{0\sigma}}
	{[i\omega-\Delta_{\sigma0}^0-V_\sigma]
		[i\omega-\Delta_{2\bar\sigma}^0]-P_{\bar\sigma2}UV_\sigma}=
\nonumber\\
=\frac{i\omega-\dote{0}-(1-\av{n_{\bar\sigma}})U}
	{[i\omega-\dote{0}-V_\sigma][i\omega-\dote{0}-U]-\av{n_{\bar\sigma}}UV_\sigma},
\label{eq-G02}
\end{eqnarray}
which is identically equal to the expression for the QD GF given in Eq. (\ref{eq-GHIA}). Hence, the QD GF expanded in terms of the Hubbard operator GFs gives exactly the same result in the HIA as for the GF given in terms of Fermi operators. This implies that the results in the non-interacting and strongly correlated limits are recovered, as well as the trivial atomic limit on the which the use of the Hubbard operators is based.

In the limit of strong on-site interaction ($U\rightarrow\infty$; doubly occupied state excluded) the result in Eq. (\ref{eq-HIAinfU}) is easily obtained in the many-body formulation, since the Hamiltonian $\Hamil_\text{int}$ can then immediately be reduced to $\Hamil_\text{int}=\sum_{p=0,\sigma}E_p\h{p}{}$ and the tunnelling term to $\Hamil_T=\sum_k(v_{k\sigma}\X{0\sigma}{}+H.c.)$. In this case it is only necessary to solve for the GF $G_{0\sigma}$, since then $G_{\bar\sigma2},\ G_{\bar\sigma2\sigma0},\ G_{0\sigma2\bar\sigma}=0$ which lead to that ${\cal G}_\sigma=G_{0\sigma}$. For $G_{0\sigma}$ one directly finds that
\[ G_{0\sigma}(i\omega)=
	\frac{P_{0\sigma}}{i\omega-\Delta_{\sigma0}^0-P_{0\sigma}V_\sigma},
\]
that is, exactly the same expression as the one given for ${\cal G}_\sigma$ in Eq. (\ref{eq-HIAinfU}).

It may be seen from the above analysis that a correct treatment of the GFs necessarily leads to self-consistent calculations of the quantities $\bfG$ and $\bfP$ involved in Eq. (\ref{eq-MHIA}) (assuming that the effects on $\bfV$ from the QD are negligible). This is clear since the end-factors $P_{0\sigma}=\Occu{(\h{0}{}+\h{\sigma}{})}=N_0+N_\sigma$ and $P_{\bar\sigma2}=\Occu{(\h{\bar\sigma}{}+\h{2}{})}=N_{\bar\sigma}+N_2$ ($U_\xi(t)\rightarrow0$), where $N_p,\ p=0,\sigma,2$, are the occupation numbers of the corresponding states $\ket{0},\ \ket{\sigma}$, and $\ket{2}$, respectively. These occupation numbers are calculated from
\begin{subequations}
\label{eq-N}
\begin{eqnarray}
N_0&=&-\frac{1}{2\pi}\im\sum_\sigma\int G_{0\sigma}^>(\omega)d\omega,
\label{eq-N0}
\\
N_\sigma&=&\frac{1}{2\pi}\im\int[G_{0\sigma}^<(\omega)
	-G_{\sigma2}^>(\omega)]d\omega,
\label{eq-N1}
\\
N_2&=&\frac{1}{2\pi}\im\sum_\sigma\int G_{\sigma2}^<(\omega)d\omega,
\label{eq-N2}
\end{eqnarray}
\end{subequations}
in non-equilibrium, where the lesser/larger form of the GFs are defined in the following section. Hence, the end-factors and GFs self-consistently depend on one another.

\subsection{Transport equation in the Hubbard I approximation}
\label{ssec-JHIA}
Clearly the GF in Eq. (\ref{eq-MHIA}) is given as a Dyson-like equation which is better seen by introducing the bare GF $\bfg$ satisfying the equation $(i\omega I-\Delta^0)\bfg(i\omega)=\bfP$. Then Eq. (\ref{eq-MHIA}) can be re-written as
\begin{equation}
\bfG=\bfg+\bfg\bfV\bfG.
\label{eq-HIAdyson}
\end{equation}
From this equation it is easy to find the retarded/advanced and lesser/larger forms of the GF, that is
\begin{eqnarray*}
\bfG^{r/a}&=&\bfg^{r/a}+\bfg^{r/a}\bfV^{r/a}\bfG^{r/a},
\\
\bfG^{</>}&=&\bfG^r\bfV^{</>}\bfG^a,
\end{eqnarray*}
where $\bfV^<=i[f_L(\omega)\bfGamma^L+f_R(\omega)\bfGamma^R]$ and $\bfV^>=-i\{[1-f_L(\omega)]\bfGamma^L+[1-f_R(\omega)]\bfGamma^R\}$, with $\bfGamma^{L/R}$ defined such that $\bfGamma^L+\bfGamma^R=\bfGamma=-2\im\bfV^r(\omega)$. The expression for $\bfG^{</>}$ is found by direct application of the Langreth rules for analytical continuation\cite{langreth1976} to the Dyson equation in Eq. (\ref{eq-HIAdyson}). The retarded/advanced and lesser GF are then inserted into the formula for the current, Eq. (\ref{eq-J}), where the trace now is taken of the $4\times4$ matrices, whereas in the case of Fermi operator representation the trace only runs over the spin indices. To show that the formula for the current coincide in the two representations, it is useful to introduce $\bfG^\text{aux}$, where the superscript stands for either of $<$ or $r/a$. The trace of $\bfGamma^\alpha\bfG^\text{aux}$, $\alpha=L,R$, is then given as
\begin{eqnarray*}
\lefteqn{
\tr\bfGamma^\alpha\bfG^\text{aux}
=}
\\&=&
	\Gamma_\up^\alpha(G_{0\up}^\text{aux}+G_{\down2\up0}^\text{aux})
	+\Gamma_\down^\alpha(G_{0\down}^\text{aux}-G_{\up2\down0}^\text{aux})
\\&&
	+\Gamma_\up^\alpha(G_{0\up2\down}^\text{aux}+G_{\down2}^\text{aux})
	+\Gamma_\down^\alpha(-G_{0\down2\up}^\text{aux}+G_{\up2}^\text{aux})=
\\&=&
	\Gamma_\up^\alpha(G_{0\up}^\text{aux}+G_{\down2\up0}^\text{aux}
	+G_{0\up2\down}^\text{aux}+G_{\down2}^\text{aux})
\\&&
	+\Gamma_\down^\alpha(G_{0\down}^\text{aux}-G_{\up2\down0}^\text{aux}
	-G_{0\down2\up}^\text{aux}+G_{\up2}^\text{aux})=
\\&=&
	\Gamma_\up^\alpha{\cal G}_\up^\text{aux}
	+\Gamma_\down^\alpha{\cal G}_\down^\text{aux}.
\end{eqnarray*}
The last line equals the trace over the spin indices of the QD GF in the Fermi operator representation. Hence, the formula for the current given in Eq. (\ref{eq-J}) is valid irrespective of whether the Fermi or Hubbard operator representation is chosen, as expected.

Here it is useful to establish a formula for the current that is valid for arbitrary $U$. From the Dyson equation of the GF $\bfG$ it is easily shown that $G_{0\sigma}^r+\eta_\sigma[G_{\bar\sigma2\sigma0}^r+G_{0\sigma2\bar\sigma}^r]+G_{\bar\sigma2}^r={\cal G}_\sigma^r$, c.f. Eq. (\ref{eq-GHIA}). Defining $V_\sigma^r=\Lambda_\sigma-i\Gamma_\sigma/2$, where $\Lambda_\sigma=\re V_\sigma^r$ and $\Gamma_\sigma^{L/R}=2\pi\sum_{k\in L/R}|v_{k\sigma}|^2\delta(\omega-\leade{k})$ such that $\Gamma_\sigma^L+\Gamma_\sigma^R=\Gamma_\sigma=-2\im V^r_\sigma$, gives the last term in Eq. (\ref{eq-J}) as 
\begin{eqnarray}
\lefteqn{
\tr[f_L(\omega)\bfGamma^L-f_R(\omega)\bfGamma^R]
		[\bfG^r(\omega)-\bfG^a(\omega)]=
}
\nonumber\\&&
	=-i\sum_\sigma
		[f_L(\omega)\Gamma_\sigma^L-f_R(\omega)\Gamma_\sigma^R]
			\Gamma_\sigma|{\cal G}_\sigma^r(\omega)|^2,
\label{eq-trGr}
\end{eqnarray}
where
\[ |{\cal G}_\sigma^r(\omega)|^2=\frac{(\omega-\Delta_{\sigma0}^0-P_{0\sigma}U)^2}
	{|(\omega-\Delta_{\sigma0}^0-V_\sigma^r)(\omega-\Delta_{2\bar\sigma}^0)
		-UP_{\bar\sigma2}V_\sigma^r|^2}.
\]
Likewise the first term of Eq. (\ref{eq-J})
\begin{eqnarray}
\lefteqn{
\tr[\bfGamma^L-\bfGamma^R]\bfG^<(\omega)=
}
\nonumber\\&&
	=i\sum(\Gamma_\sigma^L-\Gamma_\sigma^R)
		(f_L(\omega)\Gamma_\sigma^L+f_R(\omega)\Gamma_\sigma^R)
			|{\cal G}_\sigma^r(\omega)|^2,
\label{eq-trGl}
\end{eqnarray}
since $\tr\bfGamma^\alpha\bfG^r(f_L\bfGamma^L+f_R\bfGamma^R)\bfG^a=\sum_\sigma\Gamma_\sigma^\alpha(f_L\Gamma_\sigma^L+f_R\Gamma_\sigma^R)|{\cal G}_\sigma^r(\omega)|^2$. Summing the two terms in the current amounts to the formula
\begin{eqnarray}
\lefteqn{
J=\frac{e}{h}\sum_\sigma\int
	\Gamma_\sigma^L\Gamma_\sigma^R[f_L(\omega)-f_R(\omega)]
}
\nonumber\\&&
	\times\frac{(\omega-\Delta_{\sigma0}^0-P_{0\sigma}U)^2}
		{|(\omega-\Delta_{\sigma0}^0-V_\sigma^r)(\omega-\Delta_{2\bar\sigma}^0)
			-UP_{\bar\sigma2}V_\sigma^r|^2}d\omega.
\label{eq-analJ}
\end{eqnarray}
In the non-interacting limit this formula reduces to the well known result\cite{larkin1987}
\[ J=\frac{e}{h}\sum_\sigma\int
	\frac{\Gamma_\sigma^L\Gamma_\sigma^R}
		{|(\omega-\Delta_{\sigma0}^0-V_\sigma^r)|^2}
	[f_L(\omega)-f_R(\omega)]d\omega.
\]
However, also in the strongly correlated limit, $U\rightarrow\infty$, the formula for the current becomes particularly simple, e.g.
\[ J=\frac{e}{h}\sum_\sigma\int
	\frac{\Gamma_\sigma^L\Gamma_\sigma^RP_{0\sigma}^2}
		{|\omega-\Delta_{\sigma0}^0-P_{0\sigma}V_\sigma^r|^2}
	[f_L(\omega)-f_R(\omega)]d\omega,
\]
since, mathematically the only difference between the QD GF ${\cal G}_\sigma$ in the two limits is the presence of the end-factor $P_{0\sigma}$ in the strongly correlated case. Physically, the appearance of the end-factor ensures that the QD is populated by at most one electron, as expected.

\subsection{Renormalisation of the transition energies | loop correction}
\label{ssec-loop}
As was seen in Sec. \ref{ssec-mbHIA}, the the use of the Hubbard operators in the HIA may seem as an undesired complication to the problem, due to the presence of the source fields $U_\xi(t)$ and the non-trivial substitution of the three operator propagator in terms of functional derivatives of the GF. Nevertheless, the HIA served to establish a relation between the two different methods and to see that they provide equivalent results in the important limits, Eqs. (\ref{eq-atomic}), (\ref{eq-HIA0U}), and (\ref{eq-HIAinfU}). The analysis presented in this section will provide a glimpse of the power enabled within the introduced framework.

From scaling theory one finds that the localised states should be renormalised\cite{varma1976,barabanov1974,faddeev1961,haldane1978,ruckenstein1989,izyumov1992,izyumov1994} due to correlations between the QD states. Within the present approach it is possible to obtain the similar result in a rather straight forward manner and include this renormalisation of the transition energies into the GFs. The result is obtained by effecting the functional differentiation once to the GFs, neglecting the fluctuations of the end-factors. As in the HIA, all averages $\occu{\Z{\xi}{}(t)}$ but the diagonal ones are zero. The same hold for the GFs $G_{0\sigma\bar\sigma0}$ and $G_{\bar\sigma22\sigma}$ although functional derivatives thereof may not be zero, as seen below.

The structure of the equation of motion, Eq. (\ref{eq-G'}), suggests that the GF $\bfG$ is given on the form $\bfG=\bfD\bfP$, where $\bfD$ denotes the locator,\cite{sandalov2003} carrying the local on-site properties of the GF, e.g. its position and width. Hence, the variation of the GF amounts to vary both the locator and the end-factor, however, any fluctuation of the spectral weight will be omitted which then gives $\delta\bfG=(\delta\bfD)\bfP$. The locator satisfies the matrix property $\bfD\bfD^{-1}=I=\bfD^{-1}\bfD$, hence $\delta(\bfD\bfD^{-1})=(\delta\bfD)\bfD^{-1}+\bfD(\delta\bfD^{-1})=0$ which leads to the identity $\delta\bfD=-\bfD(\delta\bfD^{-1})\bfD$. Thus, it is necessary to study the locator and its inverse.

In general, the equation of motion for the locator appear very similar to Eq. (\ref{eq-G'}), however, replacing $\bfG$ by $\bfD$, and $\bfP$ by $I$ in the first term on the right hand side of Eq. (\ref{eq-G'}). The resulting equation of motion for the locator suggests that the inverted locator $\bfD^{-1}$ can be written as
\[ \bfD^{-1}(t,t')=\bfd^{-1}(t,t')-\bfS(t,t'),
\]
where the bare locator $\bfd$ satisfies the equation $[i\ddtinline-\bfDelta^0-\bfU(t)]\bfd(t,t')=\delta(t-t')I$, whereas the \emph{self-operator}\cite{sandalov2003} is identified by
\begin{eqnarray*}
\bfS(t,t')&=&\{[\bfP(t^+)+\bfR(t^+)]\kbint\bfV(t,t_1)\bfD(t_1,t_2)\}
\\&&
	\times\bfD^{-1}(t_2,t')dt_2dt_1.
\end{eqnarray*}
In the present case it is sufficient to replace the inverted locator by its corresponding bare quantity, i.e. letting $\bfD^{-1}\rightarrow\bfd^{-1}$ giving $\delta\bfD=-\bfD(\delta\bfd^{-1})\bfD$. This means that the diagrammatic expansion of the GF is terminated after the first functional differentiation, giving the so-called \emph{loop correction}.\cite{franssonPRL2002,sandalov2003,franssonPRB2002,franssonPRB2004} The above observations then lead to $\delta\bfG=-\bfD(\delta\bfd)\bfG$. Continuing the functional differentiation to higher orders generate higher order diagrams that account for additional many-body correlation effects,\cite{sandalov2003,franssonPRB2004} for instance contributions from the Kondo effect.

Application of the functional derivative, say, $R_{0\sigma\sigma'0}$ to the inverted bare locator gives $R_{0\sigma\sigma'0}(t^+)\bfd^{-1}(t_2,t_3)=-\delta(t_2-t_3)R_{0\sigma\sigma'0}(t^+)\bfU(t_2)$. It is then easy to see that
\begin{subequations}
\label{eq-RU}
\begin{eqnarray}
\lefteqn{
R_{0\sigma\sigma'0}(t^+)\bfU(t_2)=
}
\nonumber\\&&
	=-i\delta(t^+-t_2).
\label{eq-RU0}\\&&
	\times\left(\begin{array}{cccc}
	\delta_{\sigma\sigma'}-\delta_{\sigma\up}\delta_{\sigma'\up}
		& -\delta_{\sigma'\up}\delta_{\sigma\down} & 0 & 0 \\
	-\delta_{\sigma'\down}\delta_{\sigma\up}
		& \delta_{\sigma\sigma'}-\delta_{\sigma\down}\delta_{\sigma\down}
		 	& 0 & 0 \\
	0 & 0 & \delta_{\sigma\down}\delta_{\sigma'\down}
			& \delta_{\sigma'\up}\delta_{\sigma\down} \\
	0 & 0 & \delta_{\sigma'\down}\delta_{\sigma\up} 
			& \delta_{\sigma\up}\delta_{\sigma'\up}
	\end{array}\right),
\nonumber
\end{eqnarray}
\begin{eqnarray}
\lefteqn{
R_{\bar\sigma22\bar\sigma'0}(t^+)\bfU(t_2)=
}
\nonumber\\&&
	=i\delta(t^+-t_2).
\label{eq-RU2}\\&&
	\times\left(\begin{array}{cccc}
	\delta_{\bar\sigma'\up}\delta_{\bar\sigma\up}
		& \delta_{\bar\sigma\up}\delta_{\bar\sigma'\down} & 0 & 0 \\
	\delta_{\bar\sigma\down}\delta_{\bar\sigma'\up}
		& \delta_{\bar\sigma\down}\delta_{\bar\sigma'\down}
		 	& 0 & 0 \\
	0 & 0 & \delta_{\bar\sigma\bar\sigma'}
				-\delta_{\bar\sigma'\down}\delta_{\bar\sigma\down}
			& -\delta_{\bar\sigma\up}\delta_{\bar\sigma'\down} \\
	0 & 0 & -\delta_{\bar\sigma\down}\delta_{\bar\sigma'\up} 
			& \delta_{\bar\sigma\bar\sigma'}
				-\delta_{\bar\sigma\up}\delta_{\bar\sigma'\up}
	\end{array}\right),
\nonumber
\end{eqnarray}
\end{subequations}
in general.

As previously was pointed out, the GFs $G_{0\sigma\bar\sigma0}=0$ and $G_{\bar\sigma22\sigma}=0$, which implies that also $G_{0\sigma2\sigma}=0$ and $G_{\bar\sigma2\bar\sigma0}=0$. Therefore, the functional derivatives $R_{0\sigma\sigma0}$ and $R_{\bar\sigma22\bar\sigma}$ applied to the GFs give zero contribution, whereas $R_{0\sigma\bar\sigma0}$ and $R_{\bar\sigma22\sigma}$ acting on the GFs give the loop correction. Employing the results in Eqs. (\ref{eq-RU})  to the GF $G_{a\bar{b}}$ give
\begin{eqnarray}
	\left.
	\begin{array}{c}
		R_{0\sigma\bar\sigma0}(t^+)\\
		R_{\bar\sigma22\sigma}(t^+)
	\end{array}\right\}
		G_{a\bar{b}}(t'',t')=i\delta(t_2-t_3)\delta(t_2-t^+)
\nonumber\\
	\times
	\Bigl(
		D_{a\bar\sigma0}(t'',t^+)G_{0\sigma\bar{b}}(t^+,t')
\nonumber\\
	-D_{a2\sigma}(t'',t^+)G_{\bar\sigma2\bar{b}}(t^+,t')\Bigr).
\label{eq-loop}
\end{eqnarray}
Hence, in the limit $U_\xi(t)\rightarrow0$ the Fourier transformed components $G_{0\sigma\bar{a}}$ and $G_{\bar\sigma2\bar{a}}$ of Eq. (\ref{eq-G'}) reduce to
\begin{subequations}
\label{eq-G''}
\begin{eqnarray}
\lefteqn{
(i\omega-\Delta_{\sigma0}-P_{0\sigma}V_\sigma)
	G_{0\sigma\bar{a}}(i\omega)=P_{0\sigma\bar{a}}
}
\nonumber\\&&
	+[\eta_\sigma P_{0\sigma}V_\sigma+\eta_{\bar\sigma}\delta\Delta_{2\bar\sigma}]
	G_{\bar\sigma2\bar{a}}(i\omega),
\label{eq-G1''}\\
\lefteqn{
(i\omega-\Delta_{2\bar\sigma}-P_{\bar\sigma2}V_\sigma)
	G_{\bar\sigma2\bar{a}}(i\omega)=P_{\bar\sigma2\bar{a}}
}
\nonumber\\&&
	+[\eta_\sigma P_{\bar\sigma2}V_\sigma
		+\eta_{\bar\sigma}\delta\Delta_{\sigma0}]
	G_{0\sigma\bar{a}}(i\omega),
\label{eq-G2''}
\end{eqnarray}
\end{subequations}
with $\Delta_{\bar{a}}=\Delta_{\bar{a}}^0+\delta\Delta_{\bar{a}}$ and
\begin{subequations}
\label{eq-loop}
\begin{eqnarray}
\delta\Delta_{\sigma0}&=&\frac{1}{2\pi}\sum_{k\in L,R}|v_{k\bar\sigma}|^2\int
	\frac{f(\dote{k\bar\sigma})-f(\omega)}{\dote{k\bar\sigma}-\omega}
\nonumber\\&&
	\times\{-2\im[D_{0\bar\sigma}^r(\omega)
		+\eta_{\bar\sigma}D_{\sigma2\bar\sigma0}^r(\omega)]\}d\omega,
\label{eq-loop0}
\end{eqnarray}
\begin{eqnarray}
\delta\Delta_{2\bar\sigma}&=&-\frac{\eta_{\bar\sigma}}{2\pi}
	\sum_{k\in L,R}|v_{k\bar\sigma}|^2\int
	\frac{f(\dote{k\bar\sigma})-f(\omega)}{\dote{k\bar\sigma}-\omega}
\nonumber\\&&
	\times\{-2\im[D_{0\bar\sigma2\sigma}^r(\omega)
		+\eta_{\bar\sigma}D_{\sigma2}^r(\omega)]\}d\omega,
\label{eq-loop2}
\end{eqnarray}
\end{subequations}
where $D_{a\bar{b}}^r(\omega)$ is the retarded form of the locator $D_{a\bar{b}}(i\omega)$. The loop corrections to the transition energies arise due to kinematic interactions between particles in the different localised states which is induced by the presence of the de-localised conduction electrons. This is a characteristic feature of systems with interactions between localised and de-localised electron states.\cite{franssonPRL2002,sandalov2003} The effects of the loop correction on the transport properties of mesoscopic quantum systems (single and double QD, and for spin-dependent systems) have been discussed in Refs. \onlinecite{franssonPRL2002,franssonPRB2002,franssonPRB2004,franssonPRL2005}. Now, Eq. (\ref{eq-G''}) can be written as
\begin{eqnarray*}
G_{0\sigma}(i\omega)&=&\frac{(i\omega-\Delta_{2\bar\sigma}-P_{\bar\sigma2}V_\sigma)
							P_{0\sigma}}
	{\mbox{Denom}_\sigma(i\omega)},
\\
G_{\bar\sigma2\sigma0}(i\omega)&=&
	\frac{\eta_\sigma P_{\bar\sigma2}V_\sigma
		+\eta_{\bar\sigma}\delta\Delta_{\sigma0}}
	{i\omega-\Delta_{2\bar\sigma}-P_{\bar\sigma2}V_\sigma}
	G_{0\sigma}(i\omega),
\end{eqnarray*}
where
\begin{eqnarray}
\mbox{Denom}_\sigma(i\omega)&=&(i\omega-\Delta_{\sigma0}^0-V_\sigma)
		(i\omega-\Delta_{2\bar\sigma}-\delta\Delta_{\sigma0})
\nonumber\\&&
		-U(P_{\bar\sigma2}V_\sigma-\delta\Delta_{\sigma0}).
\label{eq-denom}
\end{eqnarray}
Similarly one derives
\begin{eqnarray*}
G_{\bar\sigma2}(i\omega)&=&\frac{(i\omega-\Delta_{\sigma0}-P_{0\sigma}V_\sigma)
							P_{\bar\sigma2}}
	{\mbox{Denom}_\sigma(i\omega)},
\\
G_{0\sigma2\bar\sigma}(i\omega)&=&
	\frac{\eta_\sigma P_{0\sigma}V_\sigma
		+\eta_{\bar\sigma}\delta\Delta_{2\bar\sigma}}
	{i\omega-\Delta_{\sigma0}-P_{0\sigma}V_\sigma}
	G_{\bar\sigma2}(i\omega).
\end{eqnarray*}
The result for the QD GF ${\cal G}_\sigma$ thus becomes (using that $P_{0\sigma}+P_{\bar\sigma2}=1$, $\eta_\sigma^2=1$, and $\eta_\sigma\eta_{\bar\sigma}=-1$)
\begin{eqnarray}
\lefteqn{
{\cal G}_\sigma(i\omega)=
}
\nonumber\\
&&=G_{0\sigma}(i\omega)+\eta_\sigma[G_{\bar\sigma2\sigma0}(i\omega)
	+G_{0\sigma2\bar\sigma}(i\omega)]+G_{\bar\sigma2}(i\omega)=
\nonumber\\
&&=\frac{i\omega-\Delta_{\sigma0}-\delta\Delta_{2\bar\sigma}
	-UP_{0\sigma}}
	{\mbox{Denom}_\sigma(i\omega)}.
\label{eq-Gloop}
\end{eqnarray}

It is necessary to check that the three basic limit result are consistent within this approximation. The atomic limit is again trivial since the renormalisation of the transition energies explicitly depend on the strength of the hybridisation $v_{k\sigma}$ between the localised and de-localised electrons in the system, c.f. Eq. (\ref{eq-loop}). Hence, $\delta\Delta_{\sigma0},\ \delta\Delta_{\bar\sigma2}\rightarrow0$ as $v_{k\sigma}\rightarrow0$, which then reduces Eq. (\ref{eq-Gloop}) to the result from the HIA, e.g. Eq. (\ref{eq-atomic}).

The non-interacting limit, $U\rightarrow0$, is straightforward to obtain. The denominator, Eq. (\ref{eq-denom}),  $\mbox{Denom}_\sigma(i\omega)\rightarrow(i\omega-\Delta_{\sigma0}^0-V_\sigma)(i\omega-\Delta_{\sigma0}^0-\delta\Delta_{\sigma0}-\delta\Delta_{2\bar\sigma})$, as $U\rightarrow0$. Hence, in this limit the expression in Eq. (\ref{eq-Gloop}) reduces to 
\begin{eqnarray*}
{\cal G}_\sigma(i\omega)&\rightarrow&
	\frac{i\omega-\Delta_{\sigma0}^0
		-\delta\Delta_{\sigma0}-\delta\Delta_{2\bar\sigma}}
	{(i\omega-\Delta_{\sigma0}^0-V_\sigma)
		(i\omega-\Delta_{\sigma0}^0
		-\delta\Delta_{\sigma0}-\delta\Delta_{2\bar\sigma})}
\\&=&
\frac{1}
	{i\omega-\Delta_{\sigma0}^0-V_\sigma},
\end{eqnarray*}
which is exactly equal to Eq. (\ref{eq-HIA0U}), as expected.

Before moving on to the third limit, it relevant to see whether the sum $\delta\Delta_{\sigma0}+\delta\Delta_{2\bar\sigma}\rightarrow0$ as $U\rightarrow0$, since the renormalisation of the transition energies has to be small in systems with weakly correlated particles. From the definition, Eq. (\ref{eq-loop}), it follows that
\begin{eqnarray*}
\delta\Delta_{\sigma0}+\delta\Delta_{2\bar\sigma}&\sim&
	D_{0\bar\sigma}^r+\eta_{\bar\sigma}D_{\sigma2\bar\sigma0}^r
		-\eta_{\bar\sigma}D_{0\bar\sigma2\sigma}^r-D_{\sigma2}^r
\\&=&
	\{(\omega-\Delta_{2\sigma}-P_{\sigma2}V_{\bar\sigma}^r)
\\&&
	+\eta_{\bar\sigma}[\eta_{\bar\sigma}P_{\sigma2}V_{\bar\sigma}^r
			+\eta_\sigma\delta\Delta_{\bar\sigma0}]
\\&&
	-\eta_{\bar\sigma}[\eta_{\bar\sigma}P_{0\bar\sigma}V_{\bar\sigma}^r
			+\eta_\sigma\delta\Delta_{2\sigma}]
\\&&
	-(\omega-\Delta_{\bar\sigma0}-P_{0\bar\sigma}V_{\bar\sigma}^r)\}
		/\mbox{Denom}_\sigma^r(\omega)
\\&=&
	\frac{\Delta_{\bar\sigma0}-\delta\Delta_{\bar\sigma}
	-\Delta_{2\sigma}+\delta\Delta_{2\sigma}}
		{\mbox{Denom}_\sigma^r(\omega)}
\\&=&\frac{\Delta_{\bar\sigma}^0-\Delta_{2\sigma}^0}
		{\mbox{Denom}_\sigma^r(\omega)}
	=-\frac{U}{\mbox{Denom}_\sigma^r(\omega)}.
\end{eqnarray*}
Hence, the sum of the renormalisation energies tends to zero as $U\rightarrow0$, as expected.

In the third limit, $U\rightarrow\infty$, one should not expect that the resulting expression of the GF equals the corresponding result from the HIA, since the HIA does not contain the renormalisation of the transition energies. The form of the resulting GF, however, should be similar as in the HIA, since the energy of the doubly occupied state tends to infinity whereas its population number $N_2\rightarrow0$. Indeed, dividing the GF in Eq. (\ref{eq-Gloop}) by $U$, the numerator of the GF becomes
\[ (i\omega-\Delta_{\sigma0}^0-\delta\Delta_{\sigma0}-\delta\Delta_{2\bar\sigma})/U
	-P_{0\sigma},
\]
whereas the denominator equals
\begin{eqnarray*}
(i\omega-\Delta_{\sigma0}^0-V_\sigma)(i\omega-\Delta_{2\bar\sigma}^0
	-\delta\Delta_{\sigma0}-\delta\Delta_{2\bar\sigma})/U
\\
	-(P_{\bar\sigma2}V_\sigma-\delta\Delta_{\sigma0}).
\end{eqnarray*}
From Eq. (\ref{eq-loop}) it follows that $\delta\Delta_{\sigma0}$ and $\delta\Delta_{2\bar\sigma}$ are finite for all $U$ and since $\Delta_{2\bar\sigma}/U\rightarrow1$, as $U\rightarrow\infty$, the final result becomes
\[ \lim_{U\rightarrow\infty}{\cal G}_\sigma(i\omega)=
	\frac{P_{0\sigma}}{i\omega-\Delta_{\sigma0}-P_{0\sigma}V_\sigma}.
\]
Formally, this result equals the one obtained in the HIA, e.g. Eq. (\ref{eq-HIAinfU}), with the replacement $\Delta_{\sigma0}^0\rightarrow\Delta_{\sigma0}$, and where\cite{franssonPRL2002,franssonPRB2004}
\begin{eqnarray*}
\Delta_{\sigma0}&=&\Delta_{\sigma0}^0
	+\frac{1}{2\pi}\sum_{k\in L,R}|v_{k\bar\sigma}|^2
\\&&\vphantom{\int}
	\times\int\frac{f(\dote{k\bar\sigma})-f(\omega)}
			{\dote{k\bar\sigma}-\omega}
			[-2\im D_{0\bar\sigma}^r(\omega)]d\omega.
\end{eqnarray*}
In this limit it is fairly simple to find an analytical expression for the population number $\av{n_\sigma}$, since $N_2\rightarrow0$ giving $\av{n_\sigma}=P_{\sigma2}=N_\sigma+N_2\rightarrow N_\sigma$. From Eq. (\ref{eq-N1}) one then obatins (at $T=0$ K)
\[ \av{n_\sigma}=\frac{P_{0\sigma}}{\Gamma_\sigma}
	\sum_{\alpha=L,R}\Gamma_\sigma^\alpha
	\biggl\{\frac{1}{\pi}
		\arctan{\frac{\mu_\alpha-\Delta_{\sigma0}}
			{P_{0\sigma}\Gamma_\sigma/2}}+\frac{1}{2}\biggr\}.
\]
This GF ($U\rightarrow\infty$) was derived in Ref. \onlinecite{franssonPRL2002} and is consistent with the result in Ref. \onlinecite{barabanov1974}. Letting $\Delta_{\sigma0}\rightarrow\Delta_{\sigma0}^0$ in the GF and population number, shows that these equations are consistent with the equilibrium result by Varma and Yafet,\cite{varma1976} as expected.

As in the HIA, the equations for the GFs and the end-factors have to be self-consistently solved for each bias voltage in order to ensure an accurate non-equilibrium treatment of the system. In addition, the renormalised transition energies have to be found from self-consistent calculations, since for instance the transition energy $\Delta_{\sigma0}$ depend on the all the other transition energies, through the dependence of the locators, c.f. Eq. (\ref{eq-loop}). In principle, this amounts to defining the GFs and then calculate the renormalised transition energies which should be inserted into a redefined GF, from which the occupation numbers are calculated. Self-consistency is, hence, required for the GF both with respect to the end-factors as well as the renormalised transition energies. In this sense, the presented solution with the loop correction goes far beyond the HIA, since the values of the end-factors will be influenced by the renormalised transition energies. This is further analysed in Sec. \ref{sec-numerical}.

\subsection{Transport equation with the loop correction}
\label{ssec-Jloop}
Similarly as in the HIA, the matrix equation for the Hubbard operator GFs can be set in a Dyson like equation in the same form as Eq. (\ref{eq-HIAdyson}), replacing the bare matrix $\Delta^0$ by the renormalised matrix $\Delta$. This implies that the same results for the retarded/advanced and lesser GF hold also in this case. Hence, the terms in the formula for the current, Eq. (\ref{eq-J}), are again given by Eqs. (\ref{eq-trGr}) and (\ref{eq-trGl})
while ${\cal G}_\sigma^r(\omega)$ now is given by Eq. (\ref{eq-Gloop}), replacing $i\omega$ and $V_\sigma$ by $\omega$ and $V_\sigma^r$, respectively. Summation of the two terms then gives the current expressed as
\begin{equation}
J=\frac{e}{h}\sum_\sigma\int\Gamma_\sigma^L\Gamma_\sigma^R
	[f_L(\omega)-f_R(\omega)]|{\cal G}_\sigma(\omega)|^2d\omega.
\label{eq-Jloop}
\end{equation}
In the non-interacting limit this formula reduces to the same expression as in the HIA, as expected from the discussion above, and in the strongly correlated limit ($U\rightarrow\infty$) the current becomes
\[ J=\frac{e}{h}\sum_\sigma\int
	\frac{\Gamma_\sigma^L\Gamma_\sigma^RP_{0\sigma}^2}
		{|\omega-\Delta_{\sigma0}-P_{0\sigma}V_\sigma^r|^2}
			[f_L(\omega)-f_R(\omega)]d\omega,
\]
that is, the same expression as in the HIA apart from that the transition energy here is renormalised.

\section{Charge conservation}
\label{sec-cons}
Before demonstrating the charge conservation it is relevant to show the relation $P_{0\sigma}+P_{\bar\sigma2}=1$, previously used in order to derive Eqs. (\ref{eq-G01}), (\ref{eq-G21}), and ({\ref{eq-G02}). A general proof for this relation is achieved by a direct calculation, i.e. using Eq. (\ref{eq-N}) one obtains
\begin{eqnarray*}
\lefteqn{
P_{0\sigma}+P_{\bar\sigma2}=
}
\\&&
	=\frac{1}{2\pi}\im\sum_\sigma\int[G_{0\sigma}^<-G_{0\sigma}^>
		+G_{\bar\sigma2}^<-G_{\bar\sigma2}^>]d\omega
\\&&
	=-\frac{1}{2\pi}\im\sum_\sigma\int[G_{0\sigma}^r-G_{0\sigma}^a
		+G_{\bar\sigma2}^r-G_{\bar\sigma2}^a]d\omega
\\&&
	=-\frac{1}{\pi}\im\sum_\sigma\int[G_{0\sigma}^r+G_{\bar\sigma2}^r]d\omega
\\&&=
	-\frac{1}{\pi}\tr\im\int\bfG^r(\omega)d\omega=1,
\end{eqnarray*}
since $\bfG^>-\bfG^<=\bfG^r-\bfG^a$.

Any approximate scheme used for transport calculations has to be charge conserving, and here it will be shown that both the HIA and the loop correction indeed are. In the stationary case it is sufficient to check that $\ddtinline N_p=0$. In order to show this equality, first consider the temporal derivative of the occupation number $N_0$, whereas the result for the others are obtained in a similar way. Hence,
\begin{eqnarray*}
\lefteqn{
\ddt N_0=-i\av{\com{\h{0}{}}{\Hamil}}=-i\av{\com{\h{0}{}}{\Hamil_T}}=
}
\\&=&
	-2\im\sum_{k\sigma}v_{k\sigma}^*\av{\X{\sigma0}{}\c{k}}
	=2\re\sum_{k\sigma}v_{k\sigma}^*F_{k\sigma\sigma0}^<(t,t).
\end{eqnarray*}
Using Eq. (\ref{eq-F0}) one obtains
\begin{eqnarray*}
\lefteqn{
F_{k\sigma\sigma0}^<(t,t)=
}
\\&&
	=v_{k\sigma}\int\biggl(
	g_{k\sigma}^r(t,t')[G_{0\sigma}^<(t',t)
		+\eta_\sigma G_{\bar\sigma2\sigma0}^<(t',t)]
\\&&
	+g_{k\sigma}^<(t,t')[G_{0\sigma}^a(t',t)
		+\eta_\sigma G_{\bar\sigma2\sigma0}^a(t',t)]\biggr)dt',
\end{eqnarray*}
where
\begin{eqnarray*}
g_{k\sigma}^{r/a}(t,t')&=&
	\mp i\theta(\pm t\mp t')e^{-i\leade{k}(t-t')},
\\
g_{k\sigma}^<(t,t')&=&
	if(\leade{k}) e^{-i\leade{k}(t-t')}.
\end{eqnarray*}
This gives the equation of motion for $N_0$ as
\begin{eqnarray*}
\ddt N_0&=&
	2\re\sum_{k\sigma}i|v_{k\sigma}|^2
\\&&
	\times\int_{-\infty}^t\biggl(
		f(\leade{k})[G_{0\sigma}^a(t',t)+\eta_\sigma G_{\bar\sigma2\sigma0}^a(t',t)]
\\&&
		-[G_{0\sigma}^<(t',t)+\eta_\sigma G_{\bar\sigma2\sigma0}^<(t',t)]\biggr)
		e^{-\leade{k}(t-t')}dt'
\\&=&i\sum_{k\sigma}|v_{k\sigma}|^2
\\&&
	\times\biggl(f(\leade{k})(i2\im[G_{0\sigma}^r(\leade{k})
			+\eta_\sigma G_{\bar\sigma2\sigma0}^r(\leade{k})])
\\&&
	+[G_{0\sigma}^<(\leade{k})+\eta_\sigma G_{\bar\sigma2\sigma0}^<(\leade{k})]
	\biggr).
\end{eqnarray*}
Then, using $\Gamma^{L/R}_\sigma=2\pi\sum_{k\in L/R}|v_{k\sigma}|^2\delta(\omega-\leade{k})$, one finally finds that
\begin{eqnarray*}
\lefteqn{
\ddt N_0=
}
\\&&
	\frac{i}{2\pi}\sum_{\alpha=L,R;\sigma}\Gamma_\sigma^\alpha
		\biggl(f_\alpha(\omega)(i2\im[G_{0\sigma}^r(\omega)
			+\eta_\sigma G_{\bar\sigma2\sigma0}^r(\omega)])
\\&&
	+[G_{0\sigma}^<(\omega)+\eta_\sigma G_{\bar\sigma2\sigma0}^<(\omega)]
	\biggr).
\end{eqnarray*}

By the given expressions for the GFs in the loop correction, one has that (omitting the denominators for brevity)
\begin{eqnarray*}
\lefteqn{
G_{0\sigma}^<(\omega)+\eta_\sigma G_{\bar\sigma2\sigma0}^<(\omega)\sim
}
\\&\sim&
	i(f_L(\omega)\Gamma_\sigma^L+f_R(\omega)\Gamma_\sigma^R)
		(\omega-\Delta_{2\bar\sigma}-\delta\Delta_{\sigma0})
\\&&
		\times(\omega-\Delta_{\sigma0}-\delta\Delta_{2\bar\sigma}-UP_{0\sigma})
		P_{0\sigma}
\end{eqnarray*}
and
\begin{eqnarray*}
\lefteqn{
i2\im[G_{0\sigma}^r(\omega)+\eta_\sigma G_{\bar\sigma2\sigma0}^r(\omega)]\sim
}
\\&\sim&
	-i\Gamma_\sigma
		(\omega-\Delta_{2\bar\sigma}-\delta\Delta_{\sigma0})
\\&&
	\times(\omega-\Delta_{\sigma0}-\delta\Delta_{2\bar\sigma}-UP_{0\sigma})
		P_{0\sigma}.
\end{eqnarray*}
The denominators of the two expressions are equal and, therefore, by summing over the left and right contacts one finds that
\begin{eqnarray*}
\ddt N_0\sim
	\sum_\sigma([f_L\Gamma_\sigma^L+f_R\Gamma_\sigma^R]
		\Gamma_\sigma
	-\Gamma_\sigma
		[f_L\Gamma_\sigma^L+f_R\Gamma_\sigma^R])=0,
\end{eqnarray*}
showing that the occupation number $N_0$ in the loop correction is conserved in the stationary regime. This also follows for the HIA by removing the renormalisation energies in the expressions above. Similar equalities can be shown for the other occupation numbers, e.g. $N_\sigma,\ N_2$.

In performing the analogous steps for the temporal derivatives for $N_\sigma,\ N_2$, and by summing up the occupation numbers one obtains the expression
\begin{eqnarray*}
\lefteqn{
\ddt(N_0+\sum_\sigma N_\sigma+N_2)\sim
	\tr\int\biggl(\bfGamma\bfG^<(\omega)
}
\\&&
	+[f_L(\omega)\bfGamma^L+f_R(\omega)\bfGamma^R]
	[\bfG^r(\omega)-\bfG^a(\omega)]\biggr)d\omega,
\end{eqnarray*}
which resembles the expression for the current, e.g. Eq. (\ref{eq-J}). Hence, using Eqs. (\ref{eq-trGr}) and (\ref{eq-trGl}) for the traces, replacing $\bfGamma^L-\bfGamma^R$ by $\bfGamma$ and $f_L\bfGamma^L-f_R\bfGamma^R$ by $f_L\bfGamma^L+f_R\bfGamma^R$, one easily shows that the total charge in the loop correction, as well as in the HIA, is conserved in the stationary regime. This then shows the current conservation $J_L=-J_R$, where $J_{L/R}$ is the current in the left/right lead, and the validity of the formula for the current in Eq. (\ref{eq-J}) in the given approximations.

\section{Numerical results}
\label{sec-numerical}
In this section some of the qualitative differences between the HIA and the loop correction will be analysed. The investigation is restricted to a few cases clearly showing qualitative deviations between the two approximations. A complete analysis of such differences, however, is beyond the scope of the present paper.

In both approximations, the GF along with the end-factors are self-consistently calculated for each value in the parameter space $\{\dote{0},U,\Gamma_\sigma^{L/R},V_\text{sd},k_BT\}$, where $V_\text{sd}$ is the bias voltage applied over the system. In the loop correction, the self-consistent calculations also include the renormalisation of the transition energies. The current is then calculated from knowing the QD GF, whereas the differential conductance $(dJ/dV)$ is calculated as the numerical derivative of the calculated current. This approach has been chosen since the QD GF depends on the bias voltage in a highly non-trivial way
\subsection{Non-magnetic system}
\label{ssec-nonmagn}
\begin{figure}[b]
\begin{center}
\includegraphics[width=8.5cm]{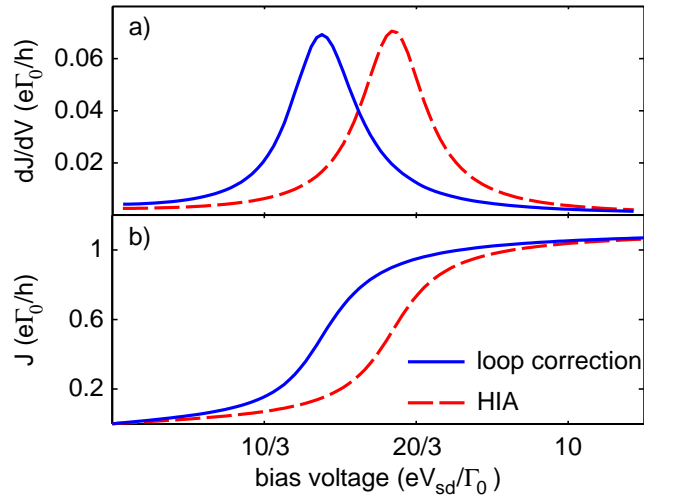}
\end{center}
\caption{(Colour online). Transport characteristics of the QD for $\{\dote{0},k_BT\}/\Gamma_0=\{3,0.014\}$ as $U\rightarrow\infty$, calculated within the HIA (dotted) and the loop correction (solid). a) The differential conductance, and b) the current, as functions of the bias voltage.}
\label{fig-jv_nm_infU}
\end{figure}
First, consider the system defined by the model, e.g.  Eqs. (\ref{eq-gmod}) and (\ref{eq-SLQD}), to be in the non-magnetic limit, i.e. $\dote{\sigma}=\dote{0}$, $\Gamma_\sigma^{L/R}=\Gamma_0/2$, and no external magnetic field. Then, as discussed, in the non-interacting limit $U\rightarrow0$ there is no difference between the HIA and the loop correction. In the opposite limit, $U\rightarrow\infty$, the difference between the HIA and loop correction is the renormalisation of the transition energies, i.e. $\delta\Delta_{\sigma0}=\Delta_{\sigma0}-\Delta_{\sigma0}^0$ which tend to lower the energy for the localised state $\ket{\sigma}$. For $\Delta_{\sigma0}^0>0\ (\Delta_{\sigma0}^0<0)$ it is expected that the current in the low bias voltage regime $|V_\text{sd}|\rightarrow0$ is smaller (larger) in the HIA than in the loop correction. This is expected since the transition $\ket{0}\bra{\sigma}$ becomes resonant for lower (higher) bias voltages in the loop correction, due to the renormalisation. Apart from this, the qualitative current-voltage ($J-V$) characteristics within the two approximations are expected to be very similar, which is verified in Fig. \ref{fig-jv_nm_infU}.

\begin{figure}[t]
\begin{center}
\includegraphics[width=8.5cm]{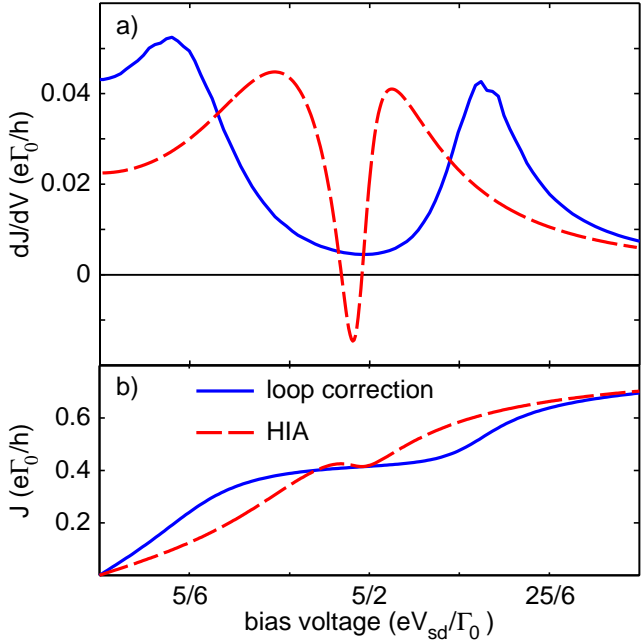}
\end{center}
\caption{(Colour online). Transport characteristics of the QD for $\{\dote{0},U,k_BT\}/\Gamma_0=\{5/6,0.5,0.014\}$, calculated within the HIA (dotted) and the loop correction (solid). a) The differential conductance, and b) the current, as functions of the bias voltage. The small ripples in the solid line in a) is due to that $dJ/dV$ is the numerical derivative of the calculated current.}
\label{fig-jv_nm_U}
\end{figure}
\begin{figure}[t]
\begin{center}
\includegraphics[width=8.5cm]{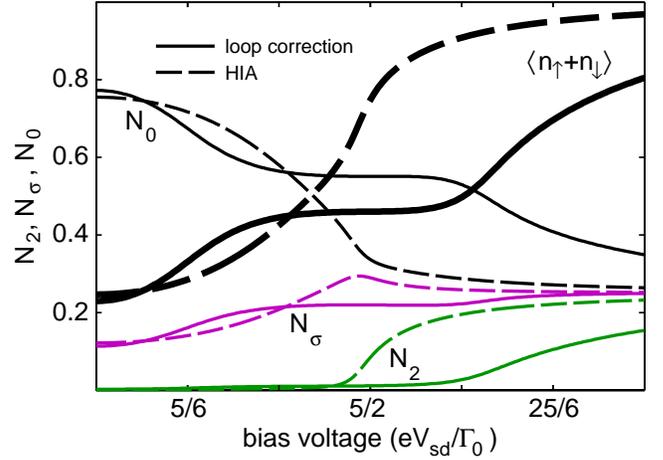}
\end{center}
\caption{(Colour online). Population numbers for the states $N_p,\ p=0,\sigma,2$ ($N_\up=N_\down$) (faint), and average population $\av{n_\up+n_\down}$ (bold) of the QD, as functions of the bias voltage calculated within the HIA (dotted) and loop correction (solid). Parameters the same as in Fig. \ref{fig-jv_nm_U}.}
\label{fig-N_nm_U}
\end{figure}
In contrast, tuning the system into the regime $1<\Delta_{\sigma0}^0/\Gamma_0<\Delta_{2\bar\sigma}^0/\Gamma_0$ (or $\Delta_{\sigma0}^0/\Gamma_0<\Delta_{2\bar\sigma}^0/\Gamma_0<-1$) and $0.5\lesssim U/\Gamma_0\lesssim1$, for low temperatures, one finds significant qualitative deviations in the transport characteristics between the HIA and loop correction. The current given in the HIA is peaked for bias voltages such that either of the chemical potentials $\mu_{L/R}$ lies between the two transition energies, see Fig. \ref{fig-jv_nm_U} (dashed). Consequently, there will be a region of a clear negative differential conductance (NDC) between the two conductance peaks corresponding to the transitions $\Delta_{\sigma0}^0$ and $\Delta_{2\bar\sigma}^0$. Such a behaviour of the transport characteristics, which is inconsistent with recent experimental data,\cite{persson1999,jung2005} is however not expected to occur in spin-degenerate single level systems. Rather, it is expected that there is a plateau in the bias voltage range where the one-particle state is resonant while the two-particle state is out of resonance. This character is captured within the loop correction, see Fig. \ref{fig-jv_nm_U} (solid).

The qualitative difference of the two approximations in this regime may be understood as follows. To be specific, consider the case $1<\Delta_{\sigma0}^0/\Gamma_0<\Delta_{2\bar\sigma}^0/\Gamma_0$ (whereas the case $\Delta_{\sigma0}^0/\Gamma_0<\Delta_{2\bar\sigma}^0/\Gamma_0<-1$ can be understood from analogous arguments). The renormalisation of the transition energies yields $\Delta_{\sigma0}\leq\Delta_{\sigma0}^0$ and $\Delta_{2\bar\sigma}\geq\Delta_{2\bar\sigma}^0$. Hence, in the loop correction the state $\ket{\sigma}$ begins to populate at lower bias voltages, see $N_\sigma$ in Fig. \ref{fig-N_nm_U} (solid), than in the HIA (dashed), whereas the state $\ket{2}$ remains unoccupied for higher voltages ($N_2$). (Notice, however, that the empty and one-particle states has a lower respectively higher population almost throughout the whole range of bias voltages. This does nonetheless not alter the following arguments, since it is the variation of the population numbers that give rise to the changes in the resulting current/differential conductance.) Therefore, the one-particle state almost fully saturate for voltages such that the two-particle state is out of resonance. In the HIA, on the other hand, the one- and two-particle states compete about the available population in the QD, since the transition energies lie closer to one another. This, in turn, leads to a depopulation of the empty state and an overpopulation of the one-particle state. In general, a high population number in the one-particle state in combination with a significant reduction of the population ($N_0$) in the empty state $\ket{0}$, tends to reduce the tunnelling probability of the one-particle state.\cite{franssonPRB2004} Hence, the overpopulation in $\ket{\sigma}$ affects the transport properties of the system such that the current decreases. Eventually, for bias voltages sufficiently large so that the two-particle state becomes resonant, the population in $\ket{\sigma}$ decreases to its "normal" saturation value which leads to an increasing current. In summary, correlation effects between particles in the different states of the QD  tend to remove overpopulation of the one-particle state which, in turn, removes the NDC for voltages such that $\ket{\sigma}$ and $\ket{2}$ is resonant and out of resonance, respectively.

\subsection{Ferromagnetic leads}
\label{ssec-ferromagn}
In spin-dependent systems there is a more significant difference between the HIA and the loop correction, than in the non-magnetic case. This is seen already by studying the renormalisation, Eq. (\ref{eq-loop}), from which it is clear that the transition energies $\Delta_{\sigma0},\ \Delta_{2\bar\sigma}$, non-linearly depend on the properties of the spin $\bar\sigma$ subband. Hence, by coupling ferromagnetic leads to the QD where, say, spin $\up$ is in majority, then causes a stronger renormalisation of the transitions $\Delta_{\down0},\ \Delta_{2\up}$ than what is experienced by $\Delta_{\up0},\ \Delta_{2\down}$. That is, the ferromagnetism in the leads induce a spin split of the transition energies in the QD. This fact has been analysed previously,\cite{franssonPRL2002,franssonEPL2005} for collinear and non-collinear alignment of the magnetisation directions in the leads. However, these studies focused on the large $U$ limit, whereas the present analysis concerns the difference in transport properties between the HIA and the loop correction for arbitrary $U$ with collinear ferromagnetic leads.

The effects considered here are related to the effective spin-dependent coupling parameters $\Gamma_\sigma^{L/R}$. These coupling parameters can be viewed as to account for spin-dependent tunnelling probabilities for electrons through the left/right interface and/or different density of spin $\up/\down$ electrons in the leads. However, using the effective couplings permits a qualitative analysis of the resulting transport properties without specifying the actual spin-dependence of the leads and/or the tunnelling interfaces.

The spin-dependence of the coupling $\Gamma_\sigma^\alpha,\ \alpha=L,R$, is parametrised in terms of  $p_\alpha\equiv(\Gamma_\up^\alpha-\Gamma_\down^\alpha)/(\Gamma_\up^\alpha+\Gamma_\down^\alpha)\in[-1,1]$, letting $\Gamma_{\up/\down}^\alpha=\Gamma_0(1\pm p_\alpha)/2$, where $\Gamma_0=\Gamma_\up^\alpha+\Gamma_\down^\alpha$. No essential physics is lost by this procedure, as was discussed by Martinek \etal\cite{martinek2003} In terms of the paramaters $p_\alpha$, one can study the transport properties of the system for parallel $(p_Lp_R>0)$ and anti-parallel $(p_Lp_R<0)$ magnetic alignment of the leads. Parametrizing in this way also allows to study the transport properties for symmetric $(|p_L|=|p_R|)$ and asymmetric  $(p_L\neq p_R)$ spin-dependence of the couplings.

In the present paper, the QD is coupled to a non-magnetic and a ferromagnetic lead, specified by $p_L=0$ and $p_R\neq0$. As in the non-magnetic case, the bare transition energies $\Delta_{\sigma0}^0,\ \Delta_{2\bar\sigma}^0$ are spin degenerate. The dressed transition energies, however, are spin split due to the spin-dependent couplings $\Gamma_\sigma^{L/R}$,\cite{franssonPRL2002,franssonEPL2005} as can be understood from Eq. (\ref{eq-loop}), and the discussion above.

\begin{figure}[t]
\begin{center}
\includegraphics[width=8.5cm]{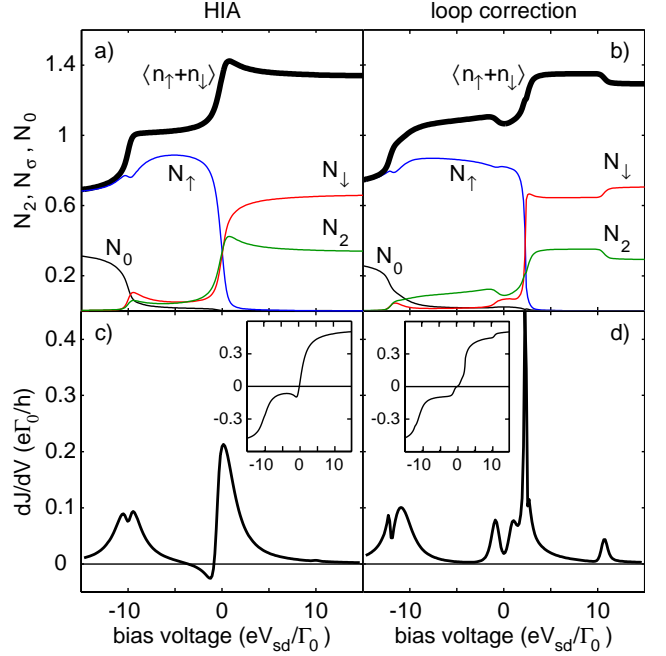}
\end{center}
\caption{(Colour online). Transport characteristics of the QD coupled to one non-magnetic lead $(p_L=0)$ and one half-metallic lead $(p_R=1)$ resulting from the HIA a) and c), and the loop correction b) and d). a), b) Population numbers $N_p,\ p=0,\sigma,2$, (faint) and $\av{n_\up+n_\down}$ (bold). c), d) Differential conductance $(dJ/dV)$. Insets show the $J-V$ characteristics of the system. Here, $\{\dote{0},U,k_BT\}/\Gamma_0=\{-5,5,0.08\}$.}
\label{fig-jv_hm}
\end{figure}
Consider the right lead being half-metallic, e.g. $p_R=1$ (or $p_R=-1$). Then from Eq. (\ref{eq-loop}), the dressed transition energies are then expected to be spin split such that $\Delta_{\down0}\leq\Delta_{\up0}$ and $\Delta_{2\down}\leq\Delta_{2\up}$ (or $\Delta_{\up0}\leq\Delta_{\down0}$ and $\Delta_{2\up}\leq\Delta_{2\down}$ ). While this behaviour of the dressed transition energies is general, the renormalisation for the transition energy $\Delta_{\sigma0}$ is stronger for large on-site Coulomb repulsion\cite{franssonPRL2002,franssonEPL2005} ($U\rightarrow\infty)$ and weakens as $U\rightarrow0$. This is also the expected property of the renormalisation from scaling theory.\cite{faddeev1961,haldane1978,ruckenstein1989,izyumov1992,izyumov1994,hewson1993} The spin split of the dressed transition energies then leads to that a current starts to flow in the spin $\up$ channel for lower bias voltages than in the spin $\down$ channel. However, in the example presented here, the will not be any current in the spin $\down$ channel since there are no available spin $\down$ states in the right lead. Hence, the total current solely consists of spin $\up$ electrons. Therefore, one cannot expect any difference in the spin polarisation of the current calculated in the HIA and in the loop correction.

Nevertheless, by tuning the system into the regime $\Delta_{\sigma0}^0/\Gamma_0\ll0$, $\Delta_{2\bar\sigma}^0/\Gamma_0\approx0$ (or $\Delta_{\sigma0}^0/\Gamma_0\approx0$, $\Delta_{2\bar\sigma}^0/\Gamma_0\gg0$), for low temperatures, one finds significant qualitative deviations in the transport characteristics between the two approximate schemes. Most interesting is that the current calculated in the HIA shows a resonant peak and associated NDC for negative biases $(eV_\text{sd}=\mu_L-\mu_R<0$) not far from equilibrium, see Fig. \ref{fig-jv_hm} c), whereas this feature is completely absent in the current and differential conductance resulting from the loop correction, Fig. \ref{fig-jv_hm} d). In these plots, the peaks around $\mp10$ and $0$, are associated with the  
transition energies $\Delta_{\up0}^*$ and $\Delta_{2\down}^*$, respectively, where $\Delta_{\bar{a}}^*$ refers to $\Delta_{\bar{a}}^0$ in the HIA and to $\Delta_{\bar{a}}$ in the loop correction. 

As in the non-magnetic case, Sec. \ref{ssec-nonmagn}, the difference in the two results can be traced down to the population numbers $N_p,\ 0,\sigma,2$, Fig. \ref{fig-jv_hm} a) and b). In both approximations, the population number $N_\down$ has finite values for $eV_\text{sd}/\Gamma_0\gtrsim-10$, and tends to zero for $eV_\text{sd}/\Gamma_0<-10$. This is plausible since the transition energy $\Delta_{\down0}^*$ lies below the chemical potential of the left lead whenever $eV_\text{sd}/\Gamma_0\gtrsim-10$, which permits a leakage of spin $\down$ electrons from the left lead into the QD. Due to this population, there is a finite probability for a double occupation of the QD, that is, that the state $\ket{2}$ acquires a population number $N_2>0$. In equilibrium, then, the population numbers $N_\up=N_\down=N_2$ in the HIA, which reflects the fact that the transition $\Delta_{2\bar\sigma}^0=\mu\ (=0)$. In the loop correction $N_\up>N_\down\approx N_2>0$, since $0<\Delta_{2\down}<\Delta_{2\up}$. As the bias voltage is turned on, such that $\mu_L-\mu_R<0$, one expects the spin $\down$ electrons to exit the QD for rather low voltages. Hence, the population number $N_\down$ is expected to approach zero, since there are no new spin $\down$ electrons entering the QD from the right. Here, the two approximations differ in the sense that $N_\down$ approaches zeros much faster in the loop correction than in the HIA, as the amplitude of the bias voltage increases. In addition, as seen in Fig. \ref{fig-jv_hm} a) and b), the population number $N_\down$ is smaller in the loop correction throughout the whole negative bias voltage range, than in the HIA. In turn, the more rapid decay and lower values of $N_\down$ in the loop correction leads to a faster increase of $N_\up$ up to its saturation value close to unity, as long as $\Delta_{\up0}$ not is resonant. Hence, when the state $\ket{\sigma}$ is almost fully occupied the current through the QD is suppressed since there is almost no weight of the state $\ket{\up}$ that can transfer electrons from the right to the left lead. However, due to the fast decay of $N_\down$, the current hardly grows larger than its value in the range where $N_\up$ is saturated. Therefore, there is no resonant peak in the current calculated in the loop correction.

Apart from the removal of the resonant peak and associated NDC, found in the HIA, there is another distinct difference between the resulting transport characteristics in the two approximations. For positive bias voltages $(\mu_L-\mu_R>0$) around $10$ mV, the transition $\Delta_{\up0}^*$ becomes resonant. Hence, the differential conductance is expected to be peaked around this voltage, Fig. \ref{fig-jv_hm} c) and d), since $\Delta_{\up0}^*$ being resonant opens a second channel for electrons to flow through, in addition to $\Delta_{2\down}^*$. Opening the transition $\ket{0}\bra{\up}$ for transport leads to a reduction of the population of spin $\up$ electrons in the QD. However, the population number $N_\up\approx0$ in almost the whole range of positive voltages, see Fig. \ref{fig-jv_hm} a) and b). The reason is that the one-particle state is fully occupied by spin $\down$, thus there cannot be any accumulation of spin $\up$ electrons in the QD since $\mu_R<\Delta_{2\down}^*<\mu_L$, meaning that any spin $\up$ electron entering the QD from the left will immediately exit the QD to the right through the transition $\ket{\down}\bra{2}$. The reduction in the population of spin $\up$ electron in the QD is possibly seen implicitly through a slight reduction of the population number $N_2$, see Fig. \ref{fig-jv_hm} a) and b). The small redistribution of the population numbers of the QD states gives rise to a small peak in the differential conductance. Nonetheless, the conductance peak is more apparent in the loop correction than in the HIA, which again is attributed to the inclusion of correlation effects in the former approximation scheme.

The double peaks in the differential conductance at around $-10$ mV in Fig. \ref{fig-jv_hm} c) and d), are due to the similar effects, that is, the transition $\ket{0}\bra{\sigma}$ becoming resonant which leads to a redistribution of the population numbers. In addition, the double peaks in the differential conductance around zero bias voltage calculated within the loop correction are due to the spin split of the transition energies, here $\Delta_{2\down}<\Delta_{2\down}$.

Finally it is worth to notice that other regimes have been considered elsewhere, for instance, the empty orbital regime $0<\Delta_{\sigma0}^0\leq\Delta_{2\bar\sigma}^0$ for similar arrangements of the spin-dependent couplings.\cite{rudzinski2001,rudzinski2004} The result found from these studies are well confirmed within the present approach (not shown here), as expected, since the present theory goes far beyond both any master equation approach or, as shown, the HIA. Nevertheless, one notices that the transport characteristics calculated within the loop corrections in general give slightly lower height of the resonant current peak (shallower NDC in the differential conductance) than what is obtained in the HIA. The arguments for this character are the same as given in the analysis of the above examples. Thus it seems as correlation effects tend to reduce, or completely remove, features like NDC in the current-voltage characteristics of the QD system.

\section{Summary}
\label{sec-sum}
An analytical formula for the current through a single level QD was derived beyond standard mean-field theory (HIA) for arbitrary on-site correlation strength. Using a diagrammatic technique based on the atomic limit properties of the interacting region, enabled an expansion of the QD GF in the strongly coupled regime. The local properties, e.g. QD GF, are solved in a self-consistent fashion with respect to the transition energies and on-site population numbers. The derived formula is consistent with previous results for single resonant level\cite{larkin1987,jauho1994} in the non-interacting limit $(U\rightarrow0)$ and with results from strongly correlated systems $(U\rightarrow\infty)$.\cite{varma1976,franssonPRL2002}

By means of the derived formula it was shown that resonant current peaks and associated NDC found in the HIA are removed by effects from electron correlations that are included into the present description (see Sec. \ref{sec-numerical}). In the non-magnetic case, the NDC in the HIA is found to arise due to an exaggerated accumulation of electron density in the QD for bias voltages such that one transition is resonant whereas the other is not.  The exaggerated population, in turn, leads to a reduced conductivity of the available transitions. However, the overestimation of the QD electron density is removed in the loop correction which then leads to a plateau in the $J-V$ characteristics, in agreement with recent experimental data.\cite{persson1999,jung2005} In the spin-dependent case, a similar overpopulation of one of the spin states $\ket{\sigma}$ in the HIA reduces the transmission through the QD. Again, this enhanced population of the QD is removed in the loop correction, whereas the resonant current peak in the $J-V$ characteristics vanish.

\acknowledgements
This work was supported by Carl Trygger Foundation, G\"oran Gustafsson's Foundation, and Swedish Foundation of Strategic Research (SSF).

\end{document}